\documentclass[usenatbib,useAMS]{mn2e}

\usepackage{graphicx}
\usepackage{bm,amssymb,amsmath}
\usepackage{times}
\usepackage{natbib}
\usepackage{float}
\usepackage{color}
\usepackage{fixltx2e} 
\usepackage{cancel}
\usepackage[normalem]{ulem} 
\graphicspath{{./fig/}{./png/}{./ps/}}

\newcommand{\vect}[1]{{{\mbox{\boldmath $#1$}}}}		

  \newcommand{\vv}{{_{\rm v}}}  				
  \newcommand{\hh}{{_{\rm h}}}  				
  \newcommand{\sound}{{_{\mathrm{s}}}}  			
  \newcommand{\dd}{\,{\rm d}}       				

  \newcommand{\m}{\,{\rm m}}
  \newcommand{\mcube}{\,{\rm m^{-3}}}

  \newcommand{\Pa}{\,{\rm Pa}}

  \newcommand{\g}{\,{\rm g}}
  \newcommand{\kg}{\,{\rm kg}}

  \newcommand{\Mm}{\,{\rm Mm}}
  \newcommand{\km}{\,{\rm km}}
  
  \newcommand{\mss}{\,{\rm m\,s^{-2}}}

  \newcommand{\K}{\,{\rm K}}

  \newcommand{\G}{\,{\rm G}}
  
  \newcommand{\mT}{\,{\rm mT}}

  \newcommand{\mkg}{\,\umu{\rm g}}

  \newcommand{\s}{\,{\rm s}}

  \newcommand{\GO}{{G}}

  \newcommand{\BO}{{B_{0z}}}

  \newcommand{\Bbz}{{B_{bz}}}

  \newcommand{\bb}{{b_{02}}}
  
  \newcommand{\bF}{{b_{01}}}
  \newcommand{\bc}{{b_{00}}}
  \newcommand{\za}{{z_{1}}}
  \newcommand{\zz}{{z_{b}}}
  
  \newcommand{\zb}{{z_{2}}}
  \newcommand{\zh}{{\ell}}
  \newcommand{\zm}{{z_{\min}}}
  \newcommand{\fO}{{f_{0}}}

  \definecolor{burntorange}{RGB}{255,97,0}
\newenvironment{freply}
  {\color{black} }

\pagerange{\pageref{firstpage}--\pageref{lastpage}}
\label{firstpage}
\pubyear{2012}
\begin{document}

  \title[Flux tube magnetohydrostatic equilibrium]{Magnetohydrostatic 
equilibrium. I: Three-dimensional open magnetic flux tube in the stratified solar
atmosphere}
  \author[Gent et al.]{
    F.~A.~Gent$^1$\thanks{
    E-mails: f.gent@shef.ac.uk,
    v.fedun@shef.ac.uk,
    s.mumford@shef.ac.uk and
    robertus@shef.ac.uk
    },
    V.~Fedun$^{2}$, S.~J.~Mumford$^1$, R.~Erd\'{e}lyi$^1$\\
    $^1$SP$^2$RC, School of Mathematics and Statistics,University of Sheffield,
    S3 7RH, UK \\
    $^2$Space Systems Laboratory, Dept. of Automatic Control and Systems
    Engineering, University of Sheffield, S1 3JD,UK      
    }
  

  \maketitle
  \begin{abstract}
    A single open magnetic flux tube spanning the solar photosphere 
    (solar radius $\simeq R_\odot$) 
    and the lower corona ($R_\odot+10\Mm$) is modelled in magnetohydrostatic 
    equilibrium within a realistic stratified atmosphere subject to solar 
    gravity. 
    Such flux tubes are observed to remain relatively stable for up to 
    a day  or more,  and it is our aim to apply the model as the background
    condition for numerical studies of energy transport mechanisms 
    from the surface to the corona. 
    We solve analytically an axially symmetric 3D structure for the model, 
    with magnetic field strength,
    plasma density, pressure and temperature all consistent with 
    observational and theoretical estimates.
    The self similar construction ensures the magnetic field is divergence
    free.
    The equation of pressure balance for this particular set of flux tubes can 
    be integrated analytically to find the pressure and density 
    corrections required to preserve the magnetohydrostatic equilibrium. 
    The model includes a number of free parameters, which makes the solution
    applicable to a variety of other physical problems and it may therefore be
    of more general interest.
    
  \end{abstract}

  \begin{keywords}
    Sun:atmosphere --- Sun: transition region  ---  instabilities --- 
    magnetic fields ---  (magnetohydrodynamics) MHD
  \end{keywords}

  \section{Introduction}\label{Intro}
  At a radius $R_\odot\simeq696\Mm$ from the Sun's core its luminous 
  surface, the photosphere, has a temperature of about $6500\K$. 
  At $h\simeq0.35-0.65\Mm$ above this surface the temperature falls to a
  minimum $T\simeq4200\K$. 
  The temperature then rises with height and experiences rapid jumps
  to $10^5\K$ just above $h\simeq2\Mm$ and to $10^6\K$ beyond 
  $h\simeq2.5\Mm$ \citep[][Ch.1, and references therein]{Priest87,Asch05}.
  The mechanism for the heating of the solar corona is not well understood. 
  The atmosphere is highly active. 
  Jets, flares, prominences and spicules carry mass and energy from the
  surface  into the atmosphere. 
  Although frequent and powerful, these solar accumulated events do not appear to
  have sufficient energy to explain the consistently high coronal 
  temperatures.
  
  Coronal loops, comprising strongly magnetized \emph{flux tubes}, 
  also permeate the
  atmosphere. 
  Given the very low thermal pressure that resides in the solar corona the magnetic
  pressure can become dynamically dominant.
  The magnetic field may be considered as a wave guide
  for carrying energy from the
  lower solar atmosphere and releasing it as heat high in the corona.
  We seek to investigate such transport mechanisms with a series of numerical 
  simulations {\freply{\citep{SFE08,FES09,SZFET09,FSE11,VFHE12}}}.
  Although transient features, these loops may persist in relative pressure
  equilibrium with the ambient atmosphere for many minutes, days or longer.
  
  In this paper a
   magnetic flux tube is modelled in pressure balance with the surrounding 
  atmosphere typical of the 
  quiet Sun.
  Modelling a realistic magnetic flux tube in magnetohydrostatic equilibrium is 
  challenging, particularly because of the exponential expansion in the radius
  of the flux tube between the photosphere and the transition region due to the
  drop in plasma pressure, and the additional constraint that the magnetic field 
  should be strong enough everywhere in the corona to provide
   the dominant pressure.
  Footpoint strengths of $100\mT~ (1000\G)$ are typically observed 
  \citep[][and references therein, Ch.8.7, Ch.5]{Zwaan78,Priest87,Asch05} and
  models with such strong fields in pressure equilibrium are often prone to 
  inducing unphysical negative thermal pressure \citep{Low80,GL98,MGRZSPTO04,GJ09}. 
  Magnetic flux tubes appear to exhibit over-dense cores in the corona 
  \citep{ASA01,WWM03}, which 
  would appear to conflict with hydrostatic equilibrium \citep{ASA01,WWM03}.
  We derive an analytic expression for a set of solutions to the 3D MHD equation
   for pressure balance with a single open magnetic flux tube.
  The physical constraints on the {\freply{plasma}} pressure, density and
  temperature are reasonably satisfied.
  
  Against this background in magnetohydrostatic equilibrium, 
  it is our intention with future work to study numerically the propagation of 
  MHD waves {\freply{through the transition region to the corona}}
  due to various physical
  drivers {\freply{in the photosphere,}} with the aim of identifying
   the primary energy transport mechanisms.
  Here we describe the analytic construction of the flux tube,
  {\freply{spanning the photosphere and about $10\Mm$ above the photosphere}}. 
  The paper is organised as follows. 
  Section~\ref{subsect:strat} details the ambient atmosphere in which the
  magnetic flux tube will be embedded, Section~\ref{subsect:mag} defines the 
  structure of the magnetic flux tube, Section~\ref{subsect:initHD} outlines
  how the atmosphere is adjusted to balance the pressure terms, 
  Section~\ref{subsect:trans} considers the necessary physical constraints 
  and in Section~\ref{sect:conc} we
  discuss the conclusions and opportunities presented by the model.
  In addition we include Appendix~\ref{sect:units}, tabulating the 
  units we use to scale the dimensionless equations, and  
  Appendix~\ref{sect:soln}, containing further details 
  of the calculations to determine the changes to the pressure and density.

  \section{The single open magnetic flux tube}\label{sect:single}
  \subsection{The stratified atmosphere}\label{subsect:strat}

  Subject to many local fluctuations, eruptions and various events on different
  scales, and varying in time depending on
  the stage of the solar cycle, the atmosphere around the solar surface may
  nevertheless be regarded as  
  predominantly in global hydrostatic equilibrium between solar gravity and 
  the total pressure gradient.
   
\begin{figure}
  \centering
  \includegraphics[width=\linewidth]{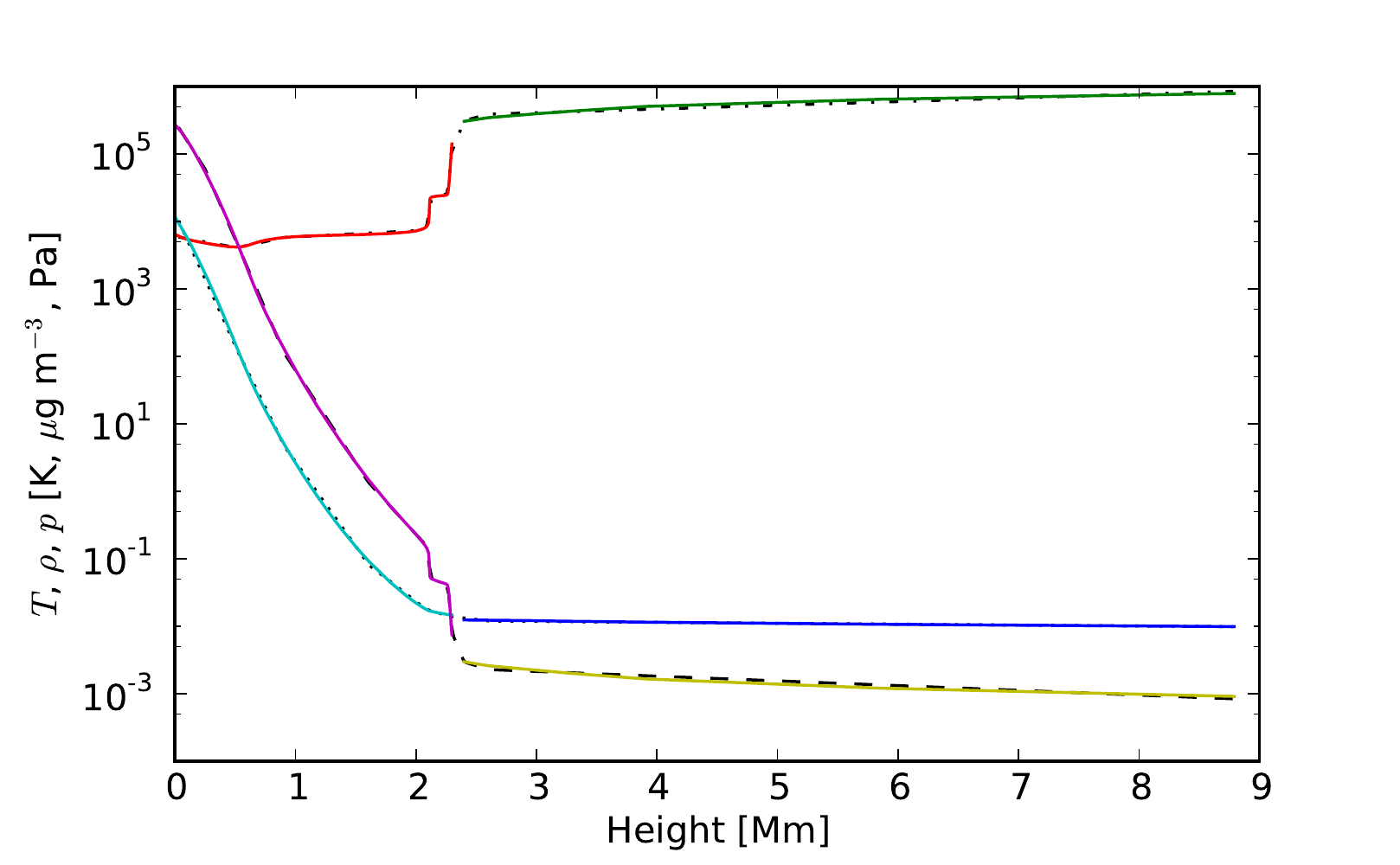}
  \caption{
  Interpolated 1D fits to vertical hydrostatic atmospheric profiles  
  \citep[][former up to 2.3\,Mm; 
  latter above 2.4\,Mm]{VAL81,MTW75}: 
  thermal pressure $p$\,[Pa] (dotted, light blue to blue),
  plasma density $\rho\,[\mkg\mcube]$ (dashed, purple to yellow) 
  and 
  temperature $T\,[\K]$ (dash-dotted, red to green). 
  \label{fig:hydroz}}
  \end{figure}

  Although accurate measurement of the atmospheric parameters is 
  challenging, due to the relatively weak intensity of the emissions from 
  the low density plasma, a number of attempts to model its structure from the 
  observational data have been recorded.
  For our model we combine the results of \citet[][Table~12,VALIIIC]{VAL81} and 
  \citet[][Table~3]{MTW75} for the chromosphere and lower solar 
  corona respectively,
  assuming parameters for the quiet Sun.
  The interpolation of these profiles as function of height above the
  surface of the photosphere are shown in Fig.~\ref{fig:hydroz}.

  In the reference data there are pronounced steps in temperature and
  density, corresponding to the transition region around $2.2\Mm$.
  The steady rise in temperature from the minimum $T\simeq4200\K$ for 
  $h\simeq500\km$ reaches the critical temperature 
  range $T>10^4\K$ over which full
  ionization of hydrogen occurs, followed subsequently by increases to the 
  critical temperatures first for single and then double ionization of the
  helium to occur almost completely. 
  To preserve the pressure equilibrium the density gradient must decrease and 
  consequently the temperature gradient also accelerate in this region until
  the plasma is almost entirely ionized.
  Thereafter temperature and density resume more steady gradients.
  The pressure gradient, however, remains relatively smooth, preserving
  the hydrostatic equilibrium.

  The pressure profiles described by \citet{VAL81} and \citet{MTW75} do not 
  include any magnetic pressure{\freply{, although a magnetic field is present and 
  therefore the total pressure is in 
  global magnetohydrostatic equilibrium.
  For our approach we require ambient conditions, in the absence of any 
  magnetic forces, to be in hydrostatic equilibrium, which these profiles are
  not. 
  We therefore need to construct such equilibrium vertical profiles from the
  reference data for density, pressure and temperature, which
  will recover the reference data profiles after we add the magnetic flux tube
  while 
  preserving magnetohydrostatic equilibrium.}}

  {\freply{The vertical pressure balance in the absence of magnetic field may be 
  expressed by
  \begin{equation}\label{eq:magp}
    \frac{\dd p\vv}{\dd z}=\rho\vv g 
   \Rightarrow \quad
     p\vv(z) = p_{\rm ref}(\zm)+\int_\zm^z \rho\vv(z^*) g \dd z^*,
  \end{equation}
  in which $p\vv$ and $\rho\vv$ represent the purely hydrothermal plasma 
  pressure and density respectively.}}
  Coordinate $\hat{\vect{z}}$ is the
  projection along the solar radial direction $\hat{\vect{R}}$ and 
  $z=0$ corresponds to $R=R_\odot$. 
  The gravitational acceleration $g$
  varies only slightly over the range of interest.
  Here it is assumed constant, $-274\mss$,
  but $g$ varying with $z$ is also applicable. 
  {\freply{$ p_{\rm ref}(\zm)\simeq10245\Pa$ is interpolated from 
  \citet{VAL81} at $\zm = 30\km$.

  From the equation of state the temperature profile is 
  \begin{equation}\label{eq:rhs}
    T\vv(z)=\frac{p\vv}{R_{\rm gas} \rho\vv},
  \end{equation}
  with the gas constant $R_{\rm gas}$. 
  The resulting pressure and temperature profiles are significiantly higher than  the reference profiles.
  An ambient average magnetic field 
  strength of up to $50\mT$ at the photosphere and $1\mT$ in the corona 
  \citep[][Ch.~1.8]{Asch05} account for the additional pressure. 
  With the magnetic field and requisite corrections to 
  plasma pressure, the reference profiles are recovered.
  To do so we also require modest 
  enhancement of the reference density profile $\rho_{\rm ref}$ to obtain 
  \begin{equation}\label{eq:ralpha}
    \rho\vv=\rho_{\rm ref}(z)+\rho_0\exp\left(-\frac{z}{z_\alpha}\right),
  \end{equation}
  with $\rho_0\simeq0.01\g\mcube$ and $z_\alpha\simeq98\km$.
  This compares to $\rho_{\rm ref}(0)\simeq0.27\g\mcube$.
  So the hydrostatic atmosphere, absent any magnetic field, is specified by 
  $p\vv,\,\rho\vv$ and $T\vv$.
 
  Here the particular choice of hydrothermal background is prescribed by the 
  solar atmosphere. 
  In general other backgrounds can be applied, subject to the requirement 
  that the pressure gradient be parallel to the flux tube.}}

  \begin{figure*}
  \centering
  \hspace{-2.25cm}
  \includegraphics[width=1.1\linewidth]{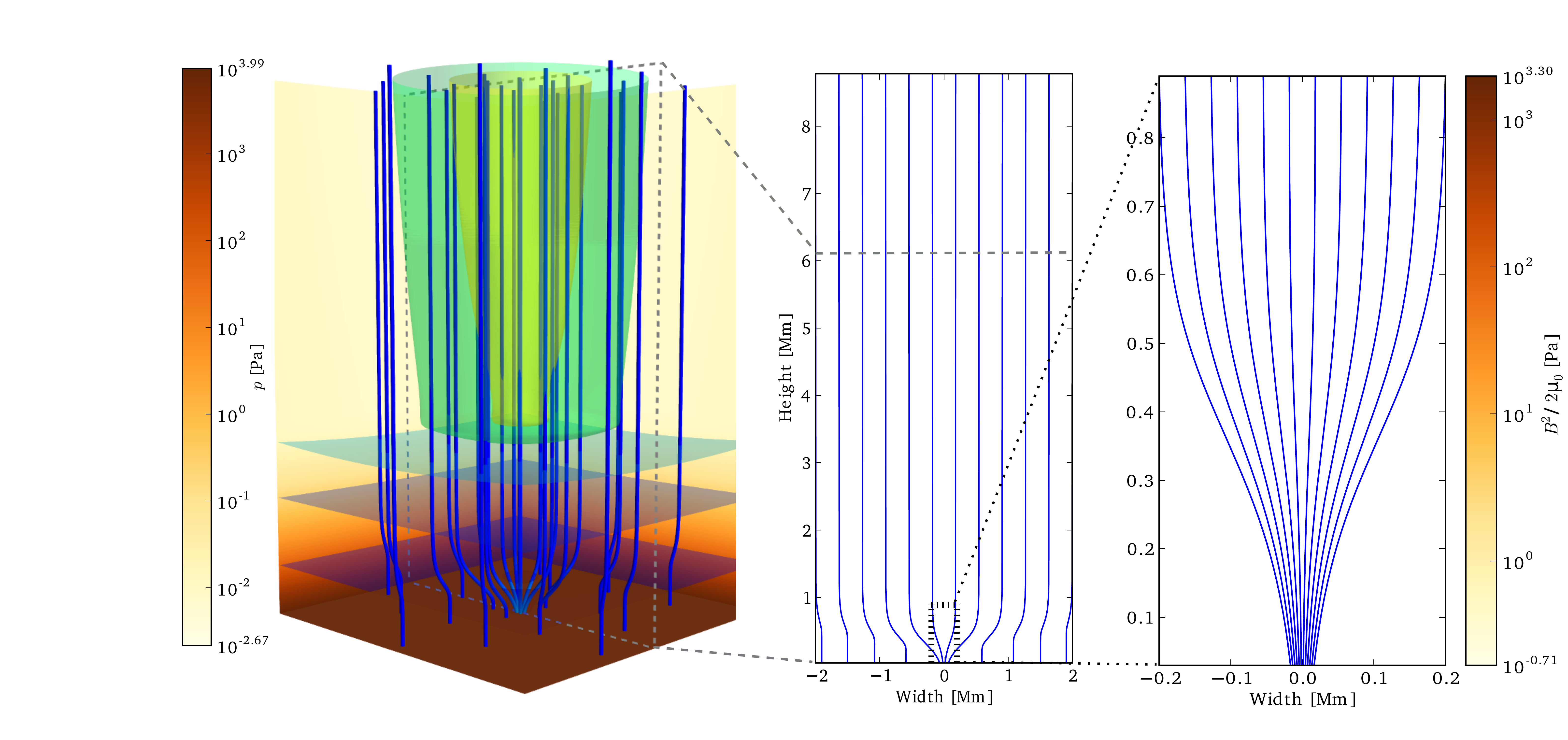}
  \caption{
  On the left a 3D rendition of the magnetic flux tube includes the magnetic 
  field lines
  (reducing field strength, turquoise -- blue). 
  The rear and bottom surfaces display the thermal pressure (reducing, 
  brown -- yellow) and the isosurfaces depict the plasma-$\beta$ 
  (purple -- green {\freply{$\simeq277, 1, 0.08, 0.025, 0.016$}}). 
  A vertical 2D-slice of the magnetohydrostatic background 
  magnetic pressure is illustrated in the middle image.
  Some representative field lines are overplotted in blue. 
  The box (black, dotted) encloses the region magnified for display in the
  image on the right.
  \label{fig:initB}}
  \end{figure*}

  \subsection{Magnetic Field Construction }\label{subsect:mag}

  {\freply{Embedded within this hydrostatic background we model a vertical open 
  magnetic flux tube, representing one footpoint of a coronal loop.
  The other footpoint is presumed to be at a distance
  beyond the horizontal extent of our numerical domain.
  The arch of the loop occurs much higher in the corona than the vertical 
  extent of our model, such that the flux tube may be regarded as
  vertically aligned. 
  The region enclosing our 
  model may reasonably be approximated either in cylindrical polar
  coordinates, with radius measured from the axis of the flux tube, or in
  Cartesian coordinates, with $x,y$ the local analogue of the longitudinal and 
  latitudinal surface coordinates.

  We elaborate the method of a self-similar expanding magnetic flux tube
  developed by \citet{ST58} and applied variously for 2D 
  \citep[e.g.][]{Deinzer65,Low80,SR05,GJ07,FSE11,SFKEM11}.

  Alternative approaches may be considered, such as the thin flux tube 
  approximation \citep[e.g.][]{RW78}. 
  To first order the effects of magnetic tension and horizontal inhomgeneity on
  the global pressure balance may be neglected.
  In our model we anticipate these effects may be significant given the 
  strong curvature of the magnetic field lines approaching the transtion region,
  and given how density inhomogeneity within each layer varies with height.
  
  Another approach is to apply a potential field to the prescribed atmosphere 
  and allow the system to relax numerically \citep[e.g.][]{SS90,KCF08}.
  Simulations of non-potential perturbations may then be applied 
  to this equilibrium.
  For models utilising very large data arrays
  there may be considerable numerical overheads before the simulations
  can proceed.
  An advantage of our approach, is that the pressure balance is specified 
  analytically, and altering the background atmosphere, perhaps to 
  represent different regions of the solar atmosphere, or to investigate 
  alternative field configurations does not require lengthy preliminary 
  numerical calculations.}}

  For a three-dimensional magnetic field describing the vertical flux tube
  {\freply{and a weak ambient field}}, we
  define its components by the relations
  {\freply{\begin{align}\label{eq:Brels}
   B_r=-\frac{\upartial f}{\upartial z}\BO\GO
       -r\frac{\upartial \Bbz}{\upartial z},~~
   B_\phi=0,~~
   B_z=\frac{\upartial f}{\upartial r}\BO\GO 
       +2\Bbz,
  \end{align}}}
  in which $\Bbz$ represents a vertically diminishing background
  term, and $\BO,\,f$ and $\GO$ prescribe the
  self-similar expanding axially symmetric  magnetic flux tube. 
  By construction $\nabla\cdot\vect{B}=0$ is preserved.
  Here $f$, $\BO$, and $\Bbz$ are defined by
  {\freply{\begin{align}\label{eq:dist}
  f=&{r}\BO\quad \quad\quad\,&[LB],\\
  \BO=&
   \bF\exp\left({-\frac{z}{\za}}\right)
     +\bb\exp\left({-\frac{z}{\zb}}\right)
 &[B],\\
  \Bbz=&\bc\exp\left(-\frac{z}{\zz}\right)&
  [B],
  \end{align}}}
  where the dimensional units for each are shown in [].
  $\bF$, $\bb$ and $\bc$ are constants, controlling the strength of the
  vertical component of the magnetic field along and around
  the axis of the flux tube. 
  $\za$ and $\zb$ are  
  included to scale the magnetic field strength along the axis with the 
  plasma pressure above and below the transition region.
  The ratio of thermal (and kinetic) to magnetic
  pressure is denoted {\emph{plasma-$\beta$}}. 
  $\zz$ scales the ambient magnetic field 
  with the pressure in the corona, 
  thus ensuring plasma-$\beta<1$ outside the flux tube and maintaining
  thermal pressure greater than zero at large $z$.

  We set  the function $\BO\GO$ to be the normalised gaussian with respect 
  to $r$ over $0\leq r<\infty$.
  The inclusion of $\BO$ in the coefficient of the gaussian is necessary to 
  ensure the shape of the flux tube is consistent as it expands to balance the
  external pressure with increasing height. 
  \begin{align}\label{eq:erf}
  \GO=&\frac{2\zh}{\sqrt{\upi}\fO}\exp\left[-
  \left(\frac{f}{\fO}\right)^2\right] \quad\quad\, &[B^{-1}].
  \end{align}
  This arrangement ensures a purely vertical magnetic
  field along the axis of the 
  flux tube and a diminishing field strength with
  increasing radius and height. 
  The argument of the gaussian function must be dimensionless so the 
  dimension of the horizontal scaling length $\fO$ is $[LB]$. 
  For the definition of the magnetic field in Eq.~\eqref{eq:Brels} to be 
  physically consistent $\GO$ must have dimension $[B^{-1}]$ and so 
  $\zh$, an appropriate normalising length scale, is
  included in the coefficient. 
  
  Explicitly the components of the magnetic field for a {flux tube} centred 
  around $r=0$ are
  {\freply{\begin{align}\label{eq:compB}
  B_r&=-f\GO\frac{\upartial \BO}{\upartial z}
     -r\frac{\upartial\Bbz}{\upartial z},\\
  B_z&={\BO^2\GO}+2\Bbz.
  \end{align}}}

  A 3D view of the flux tube is represented in the left panel 
  of Fig.~\ref{fig:initB}
  with representative magnetic field lines plotted against the
  backdrop of 
  thermal pressure and through sample isosurfaces of the plasma-$\beta$. 
  Projected from this is a vertical 2D-slice along the axis of magnetic 
  pressure overplotted with such field lines.
  These diverge radially due to the negative pressure gradient 
  below the transition
  region, but then are approximately vertical into the lower corona.  
  For closer inspection a 2D-cut near the footpoint of the flux tube
  is magnified in the right-hand panel of Fig.~\ref{fig:initB}.

  \begin{figure}
  \centering
  \includegraphics[width=\linewidth]{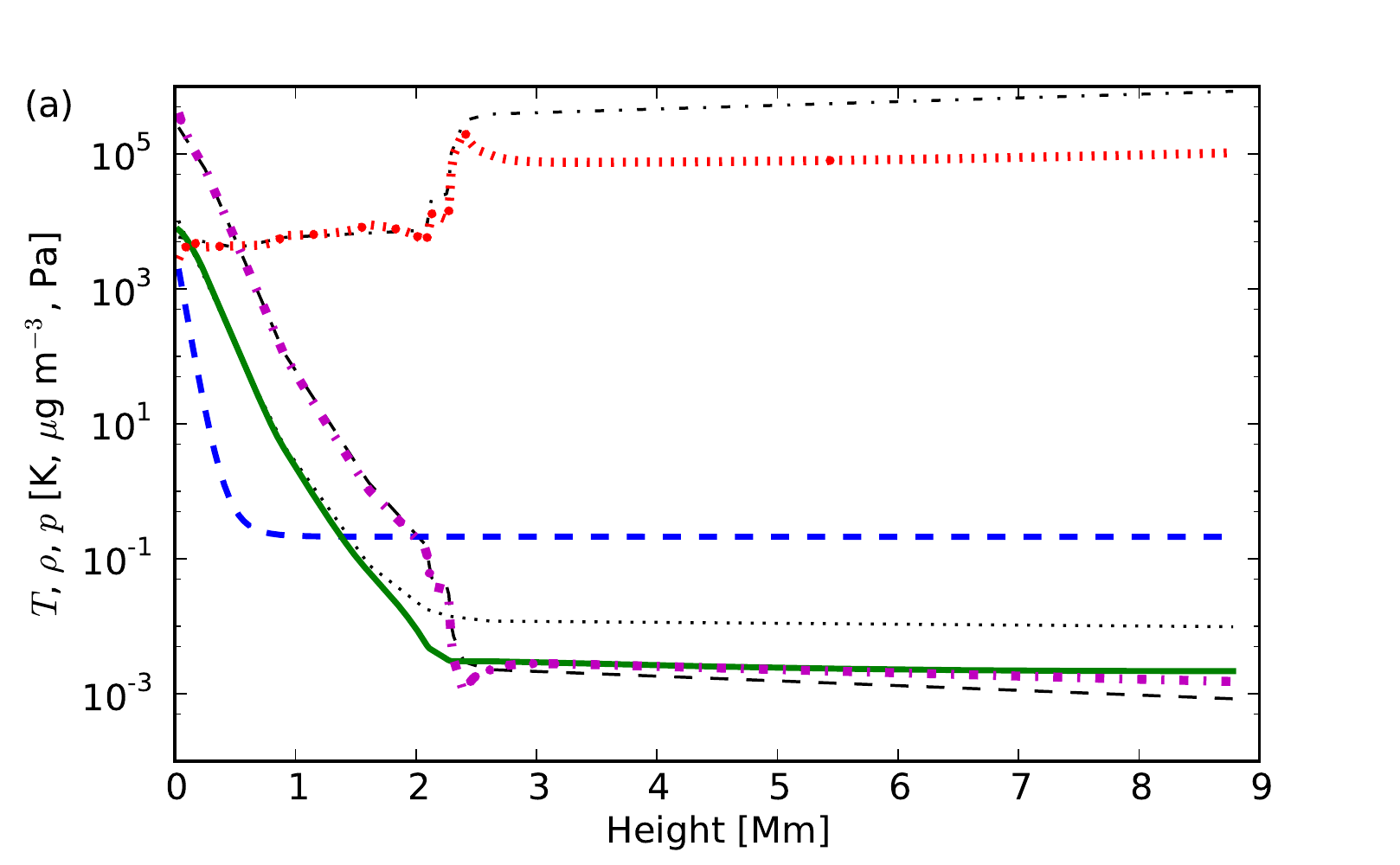}\\
  \includegraphics[width=\linewidth]{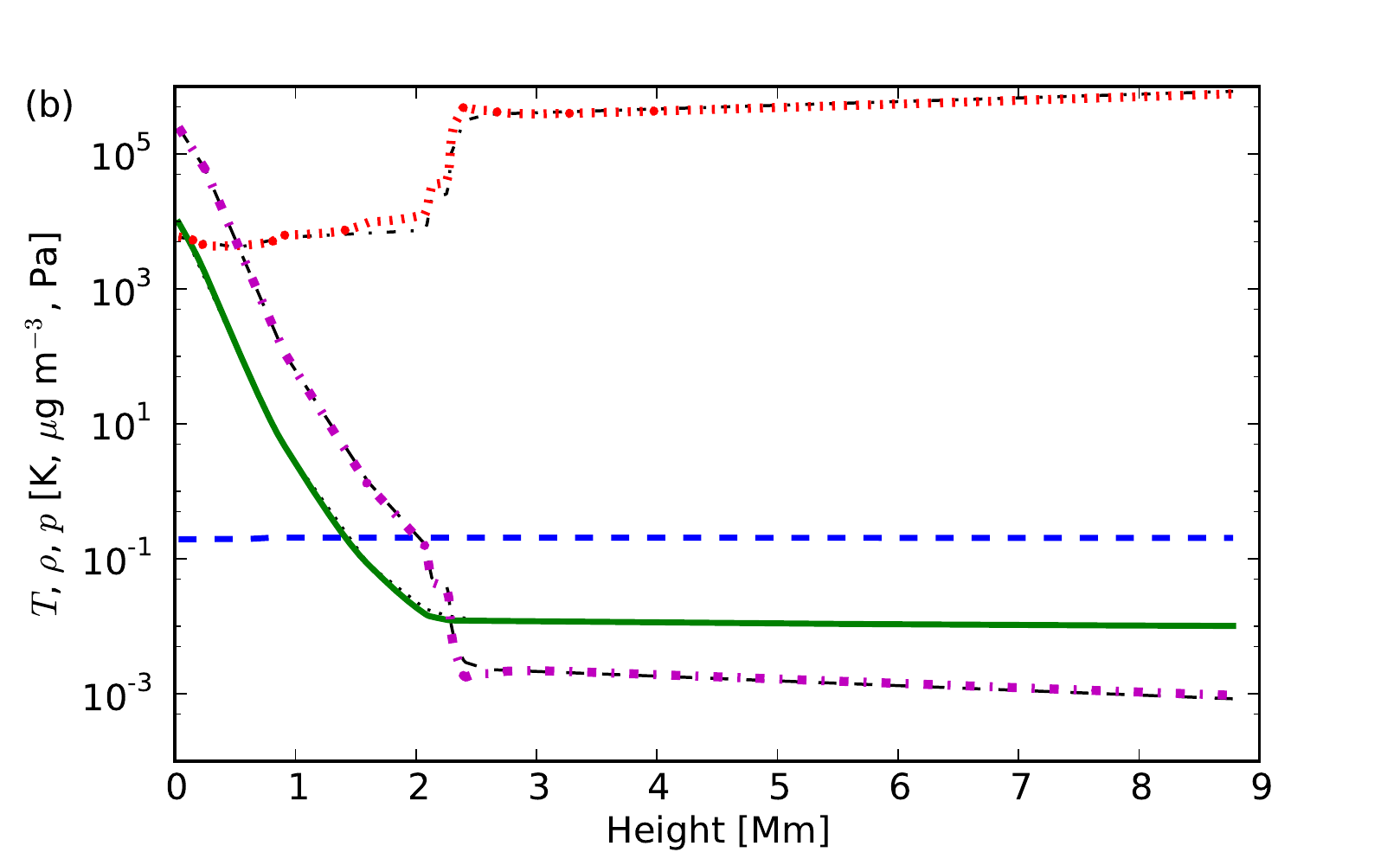}
  \caption{{\bf (a)} 
  1D-slices along the model magnetic flux tube axis of $p\,[Pa]$ 
  thermal (green, solid) and magnetic (blue, dashed) pressures, 
  $\rho\,[\mkg\mcube]$ 
  plasma density (purple, dash-dotted), 
  and  $T\,[\K]$ temperature (red, dotted)
  all superimposed on the referenced profiles (black) 
  of Fig.~\ref{fig:hydroz}.
  {\bf (b)} 
  The same 1D-slices as {\bf (a)}, but now at radius from the flux tube axis
  $r\simeq2\sqrt{2}\Mm$.
  \label{fig:initprof}}
  \end{figure}


  \begin{figure*}
  \centering
  \includegraphics[width=0.33\linewidth]{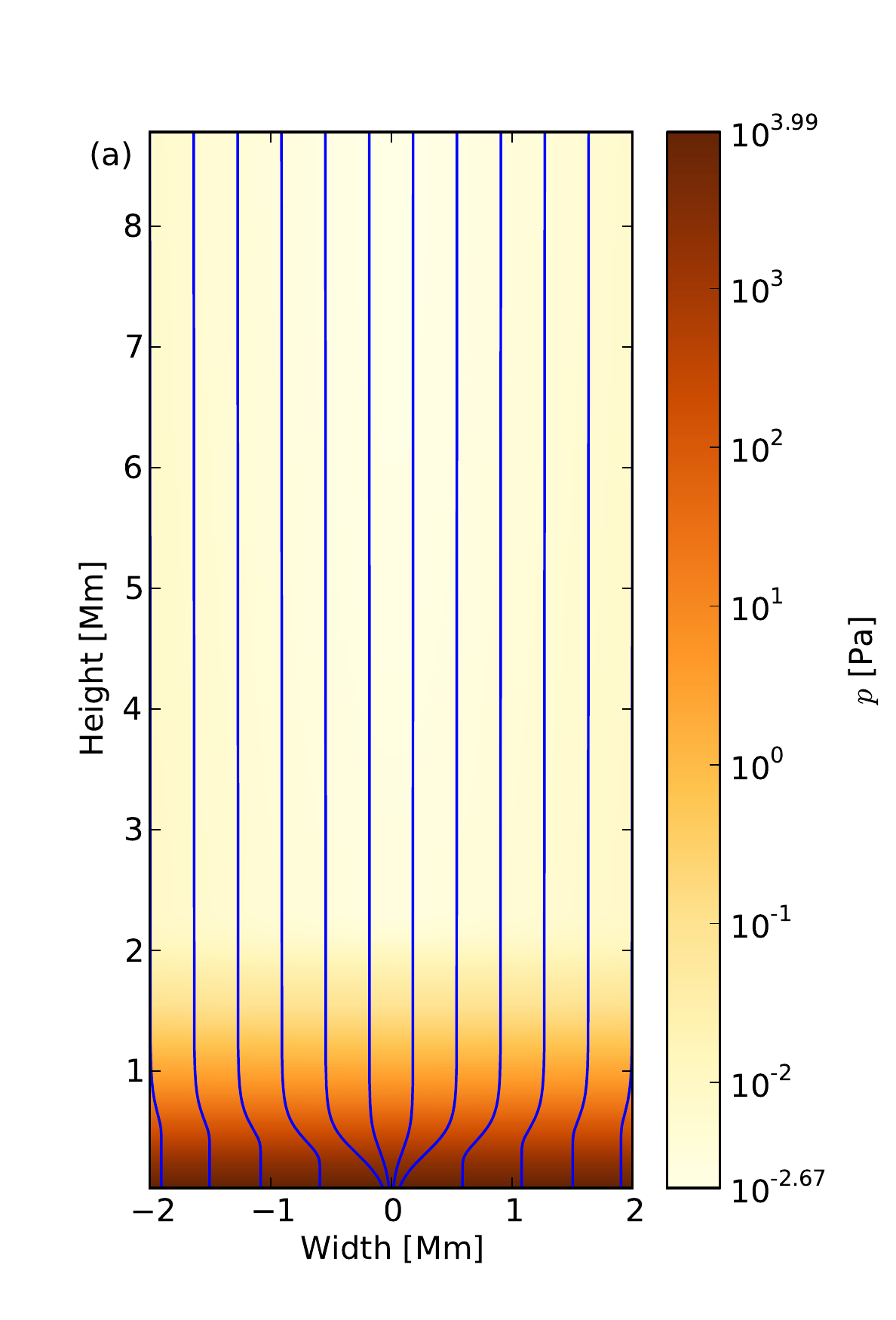}
  \includegraphics[width=0.33\linewidth]{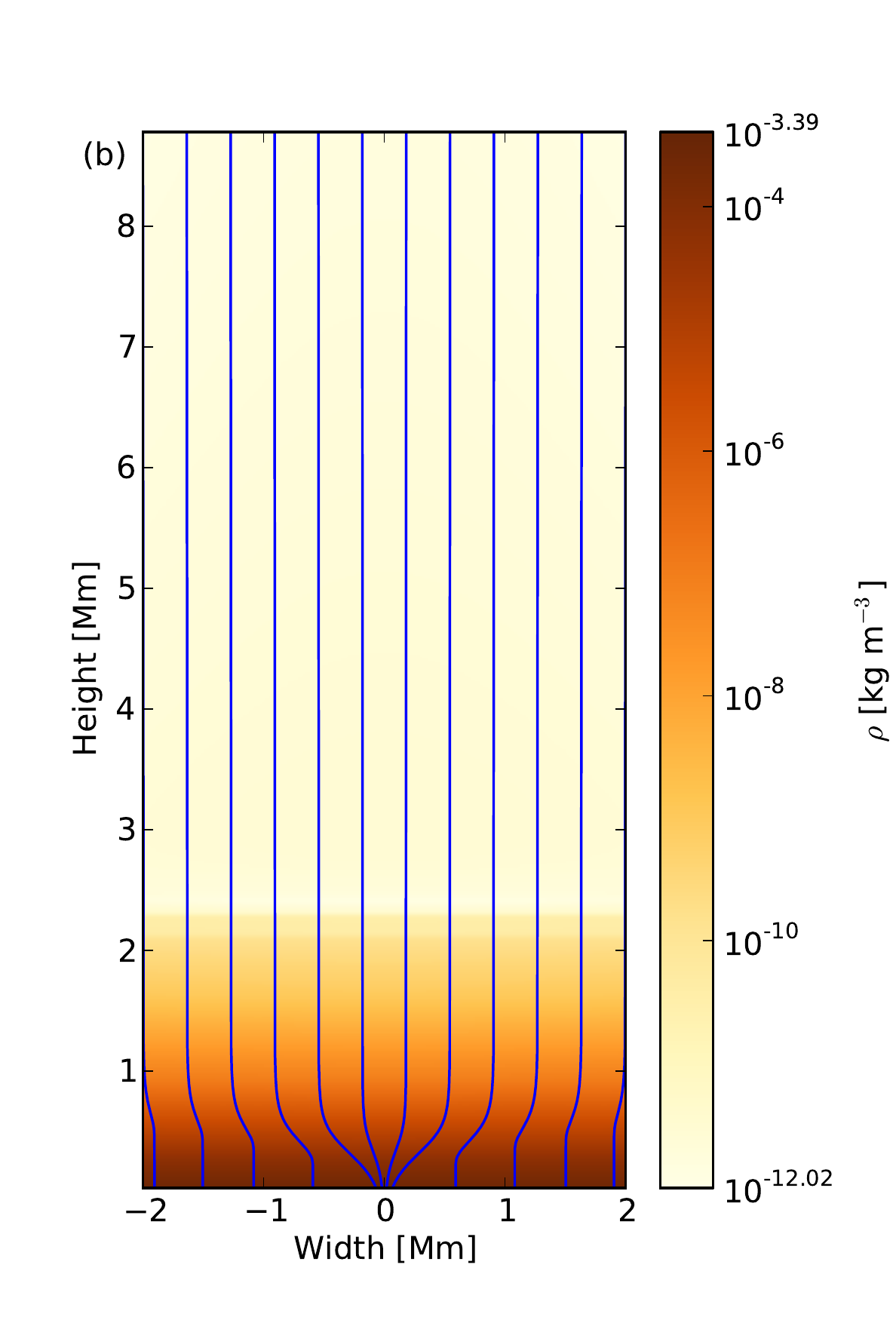}
  \includegraphics[width=0.33\linewidth]{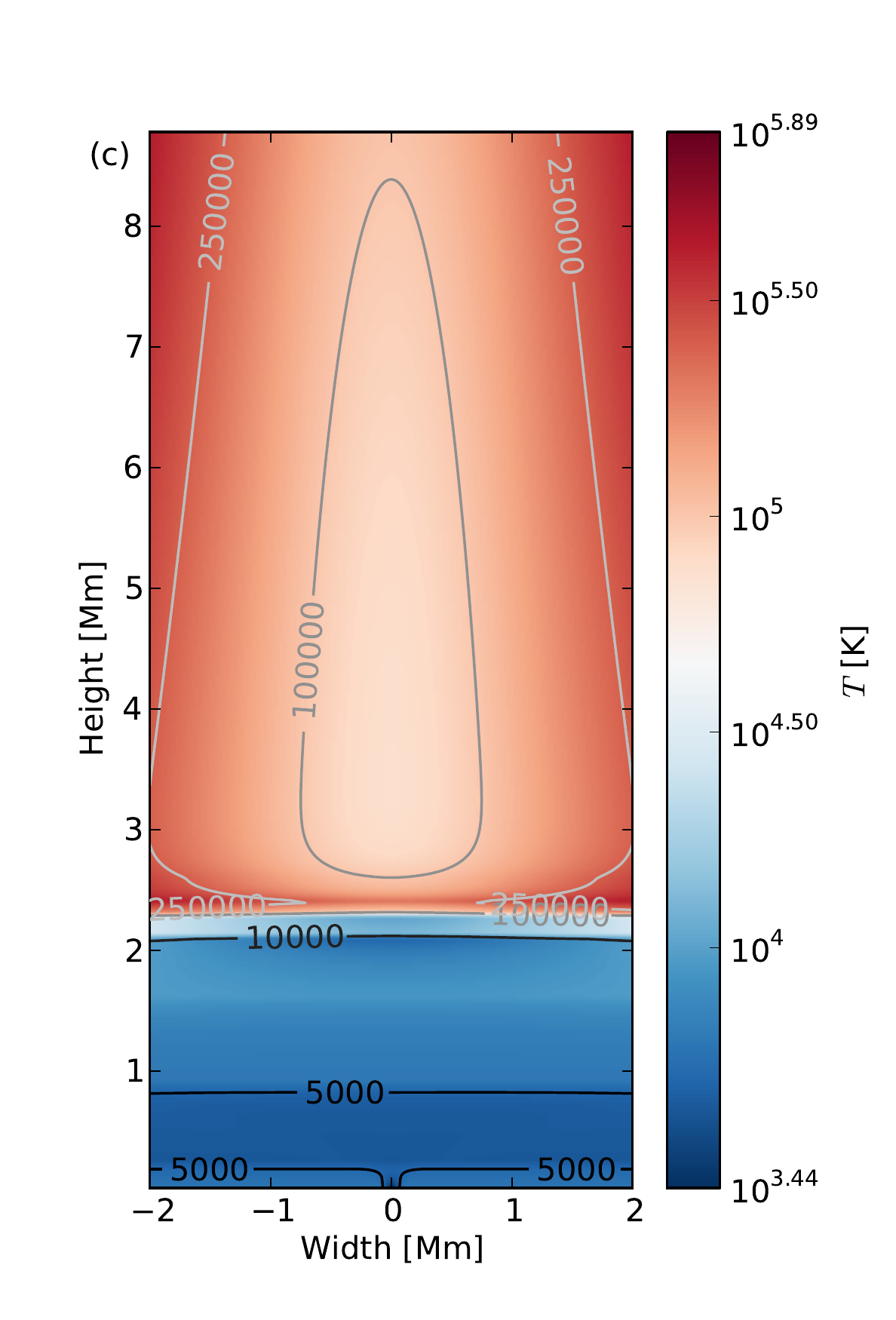}
  \caption{
  Vertical 2D-slice log profile of the magnetohydrostatic background
  {\bf (a)} thermal pressure $p$,
  {\bf (b)} 
  density $\rho$ and
  {\bf (c)} 
  temperature $T$.
  Magnetic field lines (solid, blue) are overplotted in {\bf (a)} 
  and {\bf (b)}.
  \label{fig:initp}}
  \end{figure*}

  $f$ and $\GO$ may be expressed
  in Cartesian or cylindrical polar coordinates,
  without affecting the resulting relations for pressure. 
  In Cartesian coordinates the 
  components of the magnetic field may be recast as
  {\freply{\begin{align}\label{eq:compBxy}
  B_x&=\cos\phi B_r
  =
  -x\left(\frac{\upartial\Bbz}{\upartial z}
  +{\BO\GO}
  \frac{\upartial \BO}{\upartial z}
  \right),
  \\
    B_y&=\sin\phi B_r
  =
  -y
  \left(\frac{\upartial\Bbz}{\upartial z}
  +{\BO\GO}
  \frac{\upartial \BO}{\upartial z}
  \right),
 \\
  B_z&={\BO^2\GO}+2\Bbz.
  \end{align}}}

  {\freply{Note the complexity of the magnetic field construction in this example is 
  again imposed by the structure of the lower atmosphere, incorporating the 
  transition region.
  A magnetic flux tube structure with only one exponential may be 
  adequate for modelling below the photosphere, or only
  in either the chromosphere or the corona. 
  If plasma-$\beta>1$ outside the flux is not required, the terms including
  $\Bbz$ may be neglected. 
  Conversely a more complex construction may be considered.
  Providing the terms $\BO$ and $\Bbz$ have suitable dependence only on $z$,
  the approach for finding the magnetohydrostatic corrections to $p$ 
  and $\rho$ described in this paper will apply.
  In this respect the model may have more general application.}}
\subsection{Total pressure and density} \label{subsect:initHD}

  For a background atmosphere supporting a magnetic flux tube in static 
  equilibrium the total pressure $P$ must satisfy the equation of pressure
  balance:
  \begin{equation}
  \nabla P = 
  \nabla p + \nabla \frac{|\vect{B}|^2}{2\mu_0} + (\vect{B}\cdot\nabla)
             {\freply{\frac{\vect{B}}{\mu_0} }}
           = \rho \vect{g},  \label{eq:pbeq}
  \end{equation}
  where the three inner terms are, respectively, the  
  thermal/kinetic pressure gradient,
  the magnetic pressure gradient and the magnetic 
  tension force.
  The latter is non-zero due to the curvature of the field lines. 
  $\mu_0$ is the vacuum magnetic permeability coefficient. 

  Eq.~\eqref{eq:pbeq} can be solved by integrating for each vector component
  (see
  Appendix~\ref{sect:soln} for details). 
  First, it is convenient to separate the pressure and density into parts 
  depending only on the hydrostatic pressure gradient 
  $p_\vv$ and $\rho_\vv$,
  and the horizontal corrections in the global background pressure and density 
  $p_\hh$ and $\rho_\hh$, required
  to restore the pressure balance arising from the presence of local magnetic 
  pressure and tension forces due to the magnetic flux tube.
  Thus the total pressure gradient is
  \begin{equation}\label{eq:pvh}
  {\freply{\nabla P}} = \nabla p_\vv + \nabla p_\hh +\nabla \frac{|\vect{B}|^2}{2}
           + (\vect{B}\cdot\nabla)\vect{B} = (\rho_h+\rho_v) \vect{g},
  \end{equation}
  where for convenience, the unit of magnetic field is chosen such that 
  $\mu_0=1$. 
  {\freply{$p_\vv$ and $\rho\vv$, specified by Eqs.~\eqref{eq:magp} and \eqref{eq:ralpha}
  respectively}}, are constant on the horizontal plane and independent of 
  magnetic effects, {\freply{so can be excluded from the determination of the 
  magnetohydrostatic terms.}}

  The remaining terms in Eq.~\eqref{eq:pvh} 
  are related independently of $p\vv$ and $\rho\vv$.
  The ${\vect{r}}$-component,
  \begin{equation}\label{eq:dpdr}
  \frac{\upartial p_\hh}{\upartial r} +\frac{\upartial}{\upartial r}
  \left(\frac{|\vect{B}|^2}{2}\right) + B_r\frac{\upartial B_r}{\upartial r}
   + B_z\frac{\upartial B_r}{\upartial z} = 0,
  \end{equation}
  can be integrated directly for the flux tube specified in 
  Section~\ref{subsect:mag} to obtain the thermal pressure $p\hh(r,z)$ as
  \begin{equation}\label{eq:phr}
  p_\hh=
  {\freply{B^\dagger}},
  \end{equation}
  {\freply{in which $B^\dagger$ is an expression dependent on $r,z$ as 
  detailed in 
  Eq.~\eqref{eq:phfull} of Appendix~\ref{sect:soln}}}.

  Integrating the ${\vect{z}}$-component remaining from Eq.~\eqref{eq:pvh},
\begin{equation}\label{eq:dpdz}
\frac{\upartial p_\hh}{\upartial z} +\frac{\upartial}{\upartial z}
\left(\frac{|\vect{B}|^2}{2}\right) + 
B_r\frac{\upartial B_z}{\upartial r} + B_z\frac{\upartial B_z}{\upartial z} = 
\rho\hh g,
\end{equation}
 yields a solution of the form 
\begin{equation}\label{eq:phz}
  p\hh=
{\freply{B^\dagger}}
 + \int \rho\hh g + B^*
\dd z,
\end{equation}
   in which $B^*$ 
comprises the residual terms after subtracting 
$\upartial B^\dagger/\upartial z$
from under the integral.
   Eq.~\eqref{eq:phz} must equal Eq.~\eqref{eq:phr}, requiring 
\[
\int \rho\hh g + B^*
\dd z=0.
\]
   This can be satisfied by setting 
$\rho\hh=-g^{-1}B^*$, 
  for which $B^*$ is specified in  Eq.~\eqref{eq:initr} of 
Appendix~\ref{sect:soln}.
  The thermal pressure and the density are now fully specified by
  \[
  p=p_\vv+p_\hh,\quad\rho=\rho_\vv+\rho_\hh.
  \]

  \begin{figure}
  \centering
  \includegraphics[width=\linewidth]{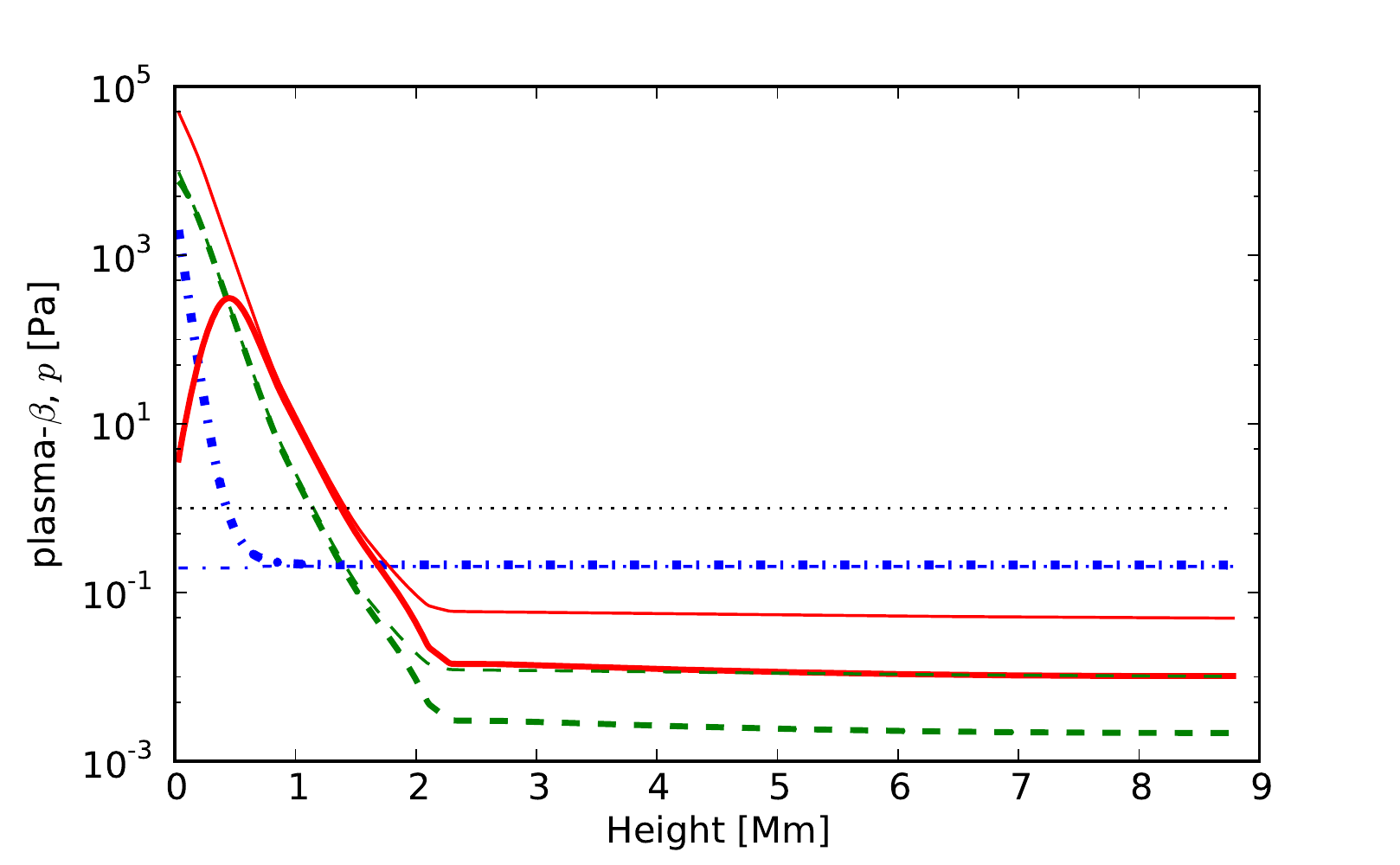}
  \caption{
  1D-slices of thermal (green, dashed) and magnetic (blue, 
  dash-dotted) pressures $p\,[Pa]$, and the plasma-$\beta$ (red, solid) along
  the magnetic flux tube axis (thick lines) and at axial radius $r\simeq2\sqrt{2}\Mm$ 
  (thin lines).
  The position of plasma-$\beta=1$ is included (black, dotted) for comparison.
  \label{fig:plasmb1}}
  \end{figure}


  The vertical profiles of the pressure, density and temperature thus 
  derived are illustrated as 1D-slices in Fig.~\ref{fig:initprof}a along 
  the axis of the magnetic flux tube and 
  in Fig.~\ref{fig:initprof}b
  outside the flux tube (at radius $r=2\sqrt2\Mm$).
  The axis of the flux tube is slightly over-dense in the corona, 
  and the temperature 
  is consequently up to an order of magnitude lower than the reference data.
  At the edge of the model the density and temperature profiles tend to
  those of the hydrostatic background.

  The vertical 2D-slices of the pressure, density and 
  temperature are also displayed in
  Fig.~\ref{fig:initp}.
  While a simulation might not extend to a radius exceeding $2\Mm$, it is 
  included here to 
  confirm that the flux tube remains physically valid beyond the numerical
  domain. 
  The model has also been checked horizontally to $\pm5\Mm$ and retains the
  features consistent with the reference data.
  {\freply{The horizontal stratification is much weaker than the vertical, so is most
  apparent in Fig.~\ref{fig:initp}c, because temperature exhibits 
  less vertical stratification than plasma pressure or density.
  The flux tube plasma is cooler than the ambient plasma. }}

  In Fig.~\ref{fig:plasmb1} the variation in plasma-$\beta$ along the 
  flux tube axis
  is plotted for the model magnetohydrostatic background along with the 
  magnetic and thermal pressure profiles.
  Note, in the corona the magnetic pressure inside and outside the flux tube is
  similar, but plasma-$\beta\lesssim0.01$ along the axis and 
  plasma-$\beta\simeq0.05$ outside differ significantly.

  \begin{figure}
  \centering
  \includegraphics[width=0.66\linewidth]{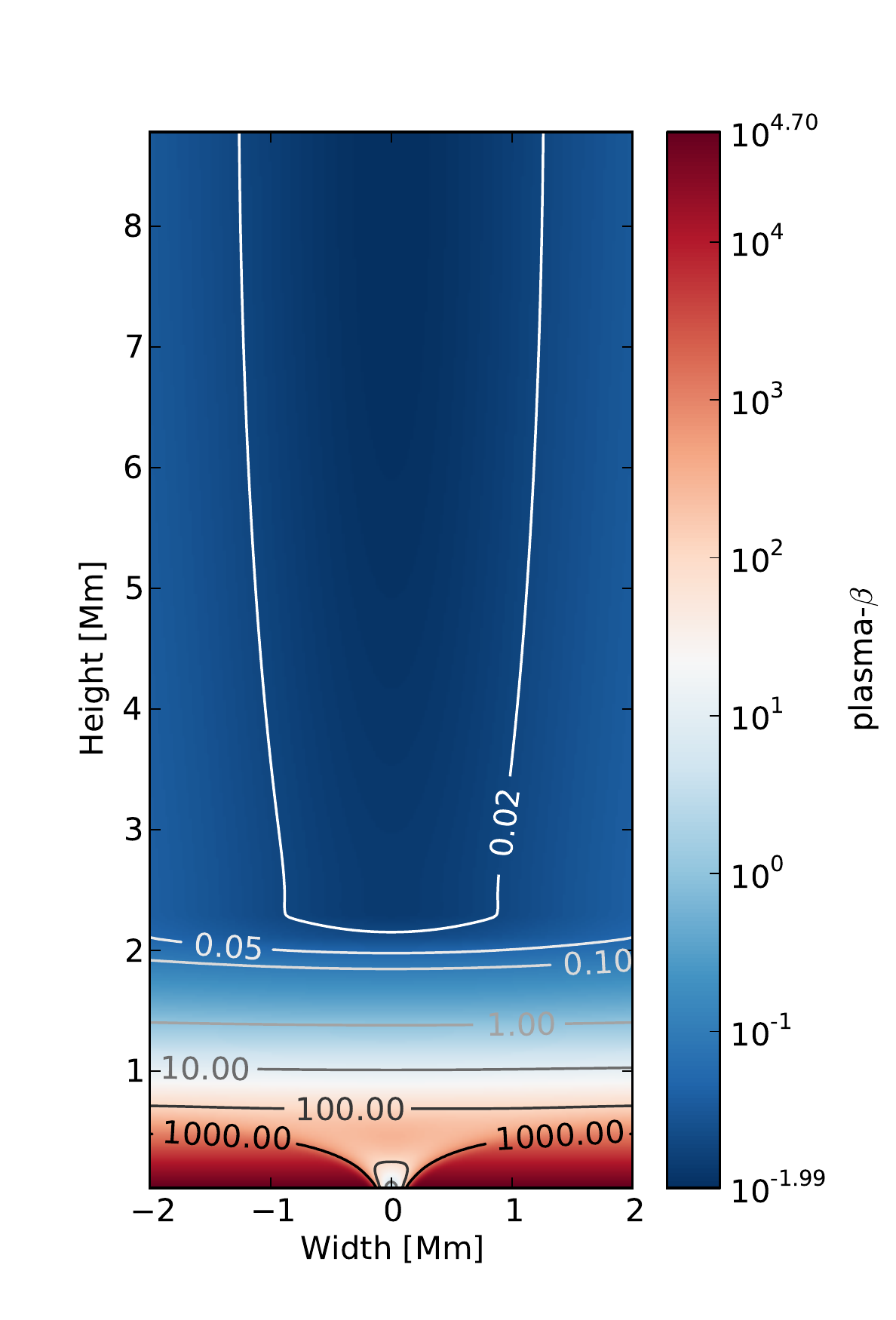}
  \caption{
  Vertical 2D-slice of the log magnetohydrostatic background plasma-$\beta$;
  the ratio of thermal to magnetic pressure. 
  \label{fig:initr}}
  \end{figure}

  The vertical 2D-slice of the log of plasma-$\beta$ is also
  depicted in Fig.~\ref{fig:initr}. 
  Note in both illustrations plasma-$\beta>1$ everywhere below $1.5\Mm$, 
  indicating the dominance of thermal pressure, and $\beta<1$ everywhere
  above, indicating the dominance of magnetic pressure even below the 
  transition region.
  There is a pronounced kink in the structure of
  the plasma-$\beta$ about $z=2.2\Mm$, corresponding to the step in plasma 
  density and temperature at the transition region. 
  Inclusion of these features may help to identify critical transport
  processes in simulations as propagating waves reach the transition region.

  \begin{figure}
  \centering
  \includegraphics[width=\linewidth]{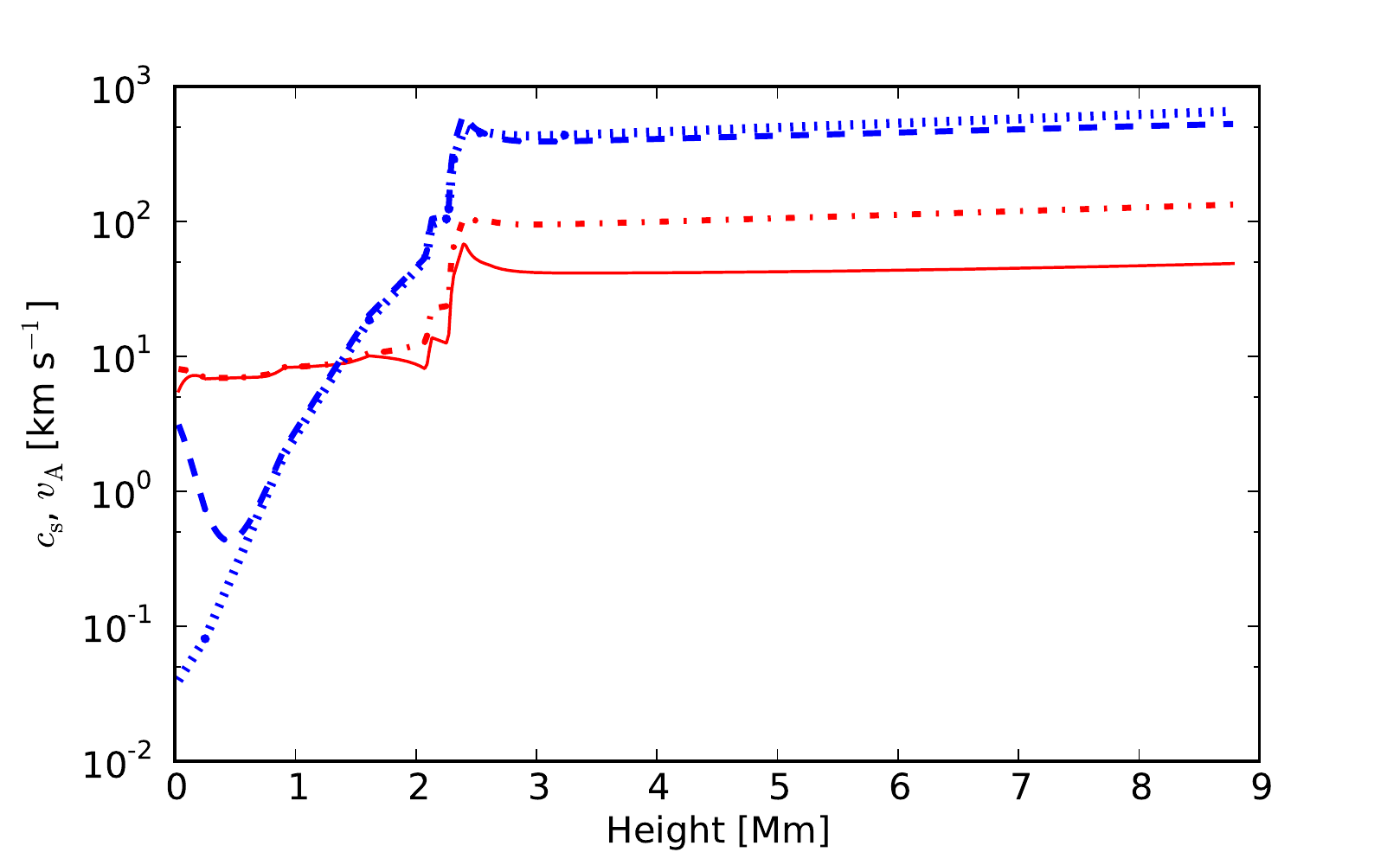}
  \caption{Vertical 1D-slices of the magnetohydrostatic background 
  sound speed $c\sound$ (red) and 
  the Alfv\'en speed $v_{\rm A}$ (blue). 
  Profiles are plotted along the magnetic flux tube axis (solid, dashed) and at axial
  radius $r\simeq2\sqrt{2}\Mm$ (dash-dotted, dotted).
  \label{fig:speed}}
  \end{figure}

  The 1D-slices of the sound speed $c\sound$ and 
  Alfv\'en speed $v_{\rm A}$ of the magnetohydrostatic background
  are displayed in 
  Fig.~\ref{fig:speed}.
  Inside and outside the magnetic flux tube $c\sound$ is similar below the
  transition region, but diverges significantly above. 
  $v_{\rm A}$ inside and outside the flux tube is quite 
  different below the temperature minimum at $z\simeq500\km$ but then is 
  similar after that. 
  In the transition region the stepped gradients of $c\sound$ and $v_{\rm A}$ 
  are very similar to each other,   
  which may mean Alfv\'en waves could be subject to reflective effects 
  analogous to those of sound waves.
 
\subsection{Avoiding negative density and unphysical 
            effects}\label{subsect:trans}

\begin{figure}
  \centering
  \includegraphics[width=\linewidth]{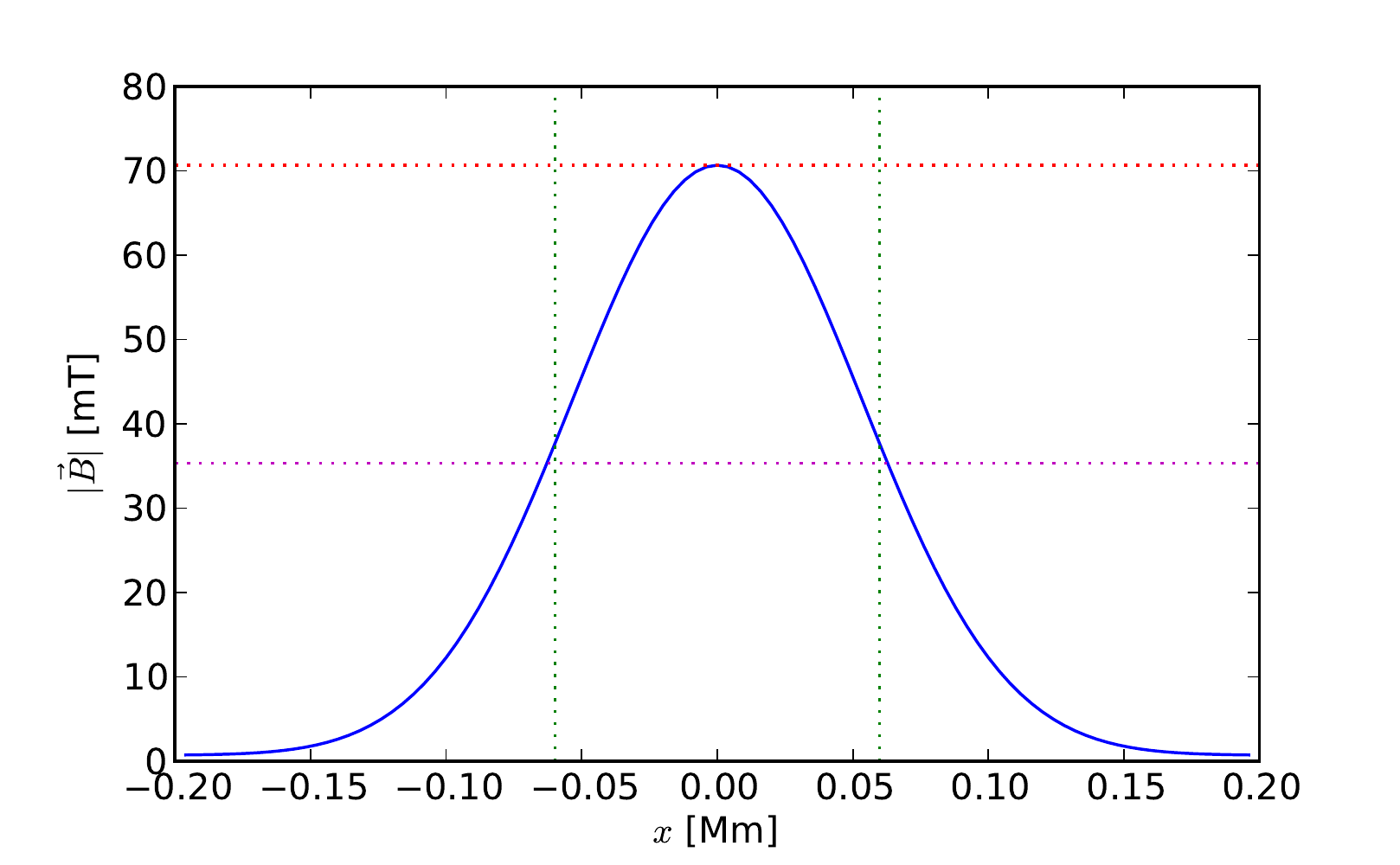}
  \caption{Horizontal 1D-slice of 
  the magnetic field strength $|\vect{B}|$ (solid, blue) through the axis of 
  the magnetic flux tube at $z=3\km$. 
  Also indicated are the FWHM, $160\km$ (vertical, dotted) and the half maximum
  (magenta, dotted).
  \label{fig:xline}}
  \end{figure}

  For our model the axial footpoint strength is $100\mT\,(1000\G)$ {\freply{at the 
  photosphere}}, yielding a full width half maximum (FWHM) of about {\freply{$100\km$}}.
  This is illustrated in Fig.~\ref{fig:xline} for $z=3\km$ with a 
  horizontal 1D-slice of the magnetic field strength {\freply{(maximum $70\mT$)}} 
  through the flux tube axis. 
  The FWHM of {\freply{$120\km$}} at $z=3\km$
  is indicated by vertical dotted lines and the half maximum by the 
  horizontal dotted line. 
  This is large enough to adequately resolve the profile with a 
  practicable numerical resolution.  

  The chosen parameters in SI units as identified in this paper
  are {\freply{$\bF\simeq0.7\mT$, $\bb\simeq0.01\mT$, $\fO\simeq40\mT\Mm$,
  $\za\simeq0.17\Mm$, $\zb\simeq175\Mm$, $\zz\simeq5\cdot10^4\Mm$
   and
  $\bc\simeq0.35\mT$.
  The scaling length $\zh\simeq8\Mm$.}}
  These parameters must be chosen to adequately track the  total pressure
  gradient, while generating a plasma-$\beta$ profile consistent with the 
  physical model. 

  Our method requires an increase in plasma density inside the magnetic flux 
  tube to balance the magnetic pressure and tension forces and so the 
  temperature is lower than outside. 
  {\freply{The mean footpoint temperature (at $z=3\km$) within a radius of $50\km$ is 
  $T\simeq3600\K$.}}
  This is low compared to observational estimates nearer to $4000\K$, however
  the model is static, while in the solar atmosphere, turbulence may effect 
  the observed temperatures and also influence the overall pressure balance.

  It is important to recognise that $p_\hh$ and $\rho_\hh$ may 
  take negative as well as positive values, subject to the constraints 
  that the sums $p_\vv+p_\hh>0$, $\rho_\vv+\rho_\hh>0$
 for any location in the domain. 
  It is also important that they are sufficiently greater than zero, such that 
  they remain positive and physically consistent even when the dynamical 
  fluctuations are included during simulations. 
  Note the thermal pressure gradient at the transition region exhibits some of 
  the stepped structure evident in the temperature and density gradients, 
  although the total pressure gradient is relatively smooth. 

  Within this transition region the plasma-$\beta$ 
   falls substantially so that magnetic pressures
  predominate. 
  This is where the density is low and rather sensitive to the strongest 
  perturbations,
  so it is essential to ensure the background $\rho$ is adequate to contain 
  any large negative perturbations.
  \section{Summary and Discussion} \label{sect:conc}
 
  We have solved analytically the MHD pressure balance equation
  for a set of single open
  vertical magnetic flux tube
  configurations in magnetohydrostatic equilibrium within a realistic 
  solar atmosphere, stratified in pressure, plasma density, temperature and 
  magnetic field strength.
  The solutions are {\freply{necessarily not inherently simplified}}, comprising a sum of 
  {\freply{multiple}} terms defining the pressure and plasma density functions, and 
  include {\freply{in this example ten}} parameters. 
  They can, however, be easily coded and visualised.
  For high performance computing the functions can also be conveniently 
  parallelized within numerical simulations.
    
  The arrangement makes it possible to include the challenging stepped 
  gradients in the 
  transition region, rather than a smooth approximation to this profile. 
  The free parameters in the model make it feasible to adjust the 
  design for numerics in order to handle strong dynamical fluctuations
  without obtaining unphysical negativity for pressure or plasma density.

  For mathematical transparency the flux tube is an idealised model,
  without torsion or any axial asymmetry, and the solar
  atmosphere is simplified
  to exclude local turbulence and fluctuations.
  However, we have endeavoured to embed the flux tube in a realistic 
  gravitationally stratified
  background atmosphere, matching closely the better estimates available from 
  theory and observation. 
  Our model does not critically depend on the prescription of the 
  ambient magnetic pressure gradient {\freply{or the precise parametarization of the 
  magnetic flux tube}}, so should data become available this 
  would constrain the model more accurately, but would not invalidate it.
  
  Exploring the magnetohydrostatic states of the model gives an indication for
  the physical constraints on magnetic field configuration, pressure, density
  and temperature, for which equilibrium is valid.  
  It appears from this result, that the over-dense features of magnetic
  flux tubes in the solar corona, may be 
  a natural prerequisite to balance the internal and external pressures.
  {\freply{With this configuration a footpoint strength in excess of $100\mT$ or 
  FWHM for this footpoint strength in excess of $100\km$ tend towards inducing 
  regions of negative plasma density or pressure.}} 
  
  Our future work 
  will include applying this analytic flux tube solution as the 
  background for numerical studies of the energy transport mechanisms between
  the photosphere and the solar corona. 
  We expect it to form the basis of a broad suite of such numerical models.
  It is worth explaining the derivation independently, which might otherwise 
  be subsumed in a more general article also relating an array of numerical
  results.   
  The aim of the present paper is to make the analytical result available for
  more general applications, further analysis and to promote the development
  of the model.
  The interactions between multiple flux tubes and alternative 
  flux tube 
  geometry might be considered, such as torsional or arched tubes.
  \section*{Acknowledgements}
  FAG is supported by STFC Grant R/131168-11-1. 
  RE is also thankful to the NSF, Hungary (OTKA, Ref. No. K83133) and 
  acknowledges M. K\'eray for patient encouragement. 
  The authors would like to acknowledge the NumPy, SciPy \citep{jones2001}, 
  Matplotlib \citep{hunter2007} and MayaVi2 \citep{ramachandran2011} python 
  projects for providing the computational tools to analyse the data.
  {\freply{We thank Andrew Soward for helpful discussion on the MHD equations, 
  Gary Verth for insight into the physical parameters of the solar atmosphere
  and Bernie Roberts on the construction of the flux tube}}. 

  \bibliographystyle{mn2e}      
  \bibliography{refs}

\begin{thebibliography}{}

\bibitem[\protect\citeauthoryear{{Aschwanden}}{{Aschwanden}}{2005}]{Asch05}
{Aschwanden} M.~J.,  2005, {Physics of the Solar Corona. An Introduction with
  Problems and Solutions (2nd edition)}

\bibitem[\protect\citeauthoryear{{Aschwanden}, {Schrijver} \&
  {Alexander}}{{Aschwanden} et~al.}{2001}]{ASA01}
{Aschwanden} M.~J.,  {Schrijver} C.~J.,    {Alexander} D.,  2001, \apj, 550,
  1036

\bibitem[\protect\citeauthoryear{{Deinzer}}{{Deinzer}}{1965}]{Deinzer65}
{Deinzer} W.,  1965, \apj, 141, 548

\bibitem[\protect\citeauthoryear{{Fedun}, {Erd{\'e}lyi} \& {Shelyag}}{{Fedun}
  et~al.}{2009}]{FES09}
{Fedun} V.,  {Erd{\'e}lyi} R.,    {Shelyag} S.,  2009, \solphys, 258, 219

\bibitem[\protect\citeauthoryear{{Fedun}, {Shelyag} \& {Erd{\'e}lyi}}{{Fedun}
  et~al.}{2011}]{FSE11}
{Fedun} V.,  {Shelyag} S.,    {Erd{\'e}lyi} R.,  2011, \apj, 727, 17

\bibitem[\protect\citeauthoryear{{Gascoyne} \& {Jain}}{{Gascoyne} \&
  {Jain}}{2009}]{GJ09}
{Gascoyne} A.,  {Jain} R.,  2009, \aap, 501, 1131

\bibitem[\protect\citeauthoryear{{Gibson} \& {Low}}{{Gibson} \&
  {Low}}{1998}]{GL98}
{Gibson} S.~E.,  {Low} B.~C.,  1998, \apj, 493, 460

\bibitem[\protect\citeauthoryear{{Gordovskyy} \& {Jain}}{{Gordovskyy} \&
  {Jain}}{2007}]{GJ07}
{Gordovskyy} M.,  {Jain} R.,  2007, \apj, 661, 586

\bibitem[\protect\citeauthoryear{Hunter}{Hunter}{2007}]{hunter2007}
Hunter J.~D.,  2007, Computing In Science \& Engineering, 9, 90

\bibitem[\protect\citeauthoryear{Jones, Oliphant, Peterson \& Others}{Jones
  et~al.}{2001}]{jones2001}
Jones E.,  Oliphant T.,  Peterson P.,    Others, 2001, {SciPy:} Open source
  scientific tools for Python

\bibitem[\protect\citeauthoryear{{Khomenko}, {Collados} \& {Felipe}}{{Khomenko}
  et~al.}{2008}]{KCF08}
{Khomenko} E.,  {Collados} M.,    {Felipe} T.,  2008, \solphys, 251, 589

\bibitem[\protect\citeauthoryear{{Low}}{{Low}}{1980}]{Low80}
{Low} B.~C.,  1980, \solphys, 67, 57

\bibitem[\protect\citeauthoryear{{Manchester}, {Gombosi}, {Roussev}, {de
  Zeeuw}, {Sokolov}, {Powell}, {T{\'o}th} \& {Opher}}{{Manchester}
  et~al.}{2004}]{MGRZSPTO04}
{Manchester} W.~B.,  {Gombosi} T.~I.,  {Roussev} I.,  {de Zeeuw} D.~L.,
  {Sokolov} I.~V.,  {Powell} K.~G.,  {T{\'o}th} G.,    {Opher} M.,  2004,
  Journal of Geophysical Research (Space Physics), 109, 1102

\bibitem[\protect\citeauthoryear{{McWhirter}, {Thonemann} \&
  {Wilson}}{{McWhirter} et~al.}{1975}]{MTW75}
{McWhirter} R.~W.~P.,  {Thonemann} P.~C.,    {Wilson} R.,  1975, \aap, 40, 63

\bibitem[\protect\citeauthoryear{{Priest}}{{Priest}}{1987}]{Priest87}
{Priest} E.~R.,  1987, {Solar magneto-hydrodynamics.}

\bibitem[\protect\citeauthoryear{Ramachandran \& Varoquaux}{Ramachandran \&
  Varoquaux}{2011}]{ramachandran2011}
Ramachandran P.,  Varoquaux G.,  2011, Computing in Science \& Engineering, 13,
  40

\bibitem[\protect\citeauthoryear{{Roberts} \& {Webb}}{{Roberts} \&
  {Webb}}{1978}]{RW78}
{Roberts} B.,  {Webb} A.~R.,  1978, \solphys, 56, 5

\bibitem[\protect\citeauthoryear{{Schl{\"u}ter} \&
  {Temesv{\'a}ry}}{{Schl{\"u}ter} \& {Temesv{\'a}ry}}{1958}]{ST58}
{Schl{\"u}ter} A.,  {Temesv{\'a}ry} S.,  1958, in {Lehnert} B.,  ed.,
  Electromagnetic Phenomena in Cosmical Physics Vol.~6 of IAU Symposium, {The
  Internal Constitution of Sunspots}.
p.~263

\bibitem[\protect\citeauthoryear{{Sch{\"u}ssler} \& {Rempel}}{{Sch{\"u}ssler}
  \& {Rempel}}{2005}]{SR05}
{Sch{\"u}ssler} M.,  {Rempel} M.,  2005, \aap, 441, 337

\bibitem[\protect\citeauthoryear{{Shelyag}, {Fedun} \& {Erd{\'e}lyi}}{{Shelyag}
  et~al.}{2008}]{SFE08}
{Shelyag} S.,  {Fedun} V.,    {Erd{\'e}lyi} R.,  2008, \aap, 486, 655

\bibitem[\protect\citeauthoryear{{Shelyag}, {Fedun}, {Keenan}, {Erd{\'e}lyi} \&
  {Mathioudakis}}{{Shelyag} et~al.}{2011}]{SFKEM11}
{Shelyag} S.,  {Fedun} V.,  {Keenan} F.~P.,  {Erd{\'e}lyi} R.,
  {Mathioudakis} M.,  2011, Annales Geophysicae, 29, 883

\bibitem[\protect\citeauthoryear{{Shelyag}, {Zharkov}, {Fedun}, {Erd{\'e}lyi}
  \& {Thompson}}{{Shelyag} et~al.}{2009}]{SZFET09}
{Shelyag} S.,  {Zharkov} S.,  {Fedun} V.,  {Erd{\'e}lyi} R.,    {Thompson}
  M.~J.,  2009, \aap, 501, 735

\bibitem[\protect\citeauthoryear{{Solanki} \& {Steiner}}{{Solanki} \&
  {Steiner}}{1990}]{SS90}
{Solanki} S.~K.,  {Steiner} O.,  1990, \aap, 234, 519

\bibitem[\protect\citeauthoryear{{T\'oth}}{{T\'oth}}{1996}]{Toth96}
{T\'oth} G.,  1996, Astrophys. Lett. \& Commun., 34, 245

\bibitem[\protect\citeauthoryear{{Vernazza}, {Avrett} \& {Loeser}}{{Vernazza}
  et~al.}{1981}]{VAL81}
{Vernazza} J.~E.,  {Avrett} E.~H.,    {Loeser} R.,  1981, \apjs, 45, 635

\bibitem[\protect\citeauthoryear{{Vigeesh}, {Fedun}, {Hasan} \&
  {Erd{\'e}lyi}}{{Vigeesh} et~al.}{2012}]{VFHE12}
{Vigeesh} G.,  {Fedun} V.,  {Hasan} S.~S.,    {Erd{\'e}lyi} R.,  2012, \apj,
  755, 18

\bibitem[\protect\citeauthoryear{{Winebarger}, {Warren} \&
  {Mariska}}{{Winebarger} et~al.}{2003}]{WWM03}
{Winebarger} A.~R.,  {Warren} H.~P.,    {Mariska} J.~T.,  2003, \apj, 587, 439

\bibitem[\protect\citeauthoryear{{Zwaan}}{{Zwaan}}{1978}]{Zwaan78}
{Zwaan} C.,  1978, \solphys, 60, 213

\end{thebibliography}
  \label{lastpage}
 
\appendix
\section{Dimensional quantities}\label{sect:units}
  \begin{table}
  \caption{\label{tab:units}
  Dimensions of physical quantities used to non-dimensionalise the numerical
  equations.  
  }
{{
    \begin{tabular}{lccll}
  \hline
  Physics                 &Symbol  &Dimension                             &Units                           &              \\
  \hline                                                                                                   
  Length                  &$L$     &$[L]$                                 &$2\times10^3$       \hspace{-0.2cm}            &$\m$          \\
  Velocity                &$u$     &$[u]$                                 &$10^3$              \hspace{-0.2cm}            &$\m\s^{-1}$   \\
  Density                 &$\rho$  &$[\rho]$                              &$10^{-6}$           \hspace{-0.2cm}            &$\kg\mcube$   \\
  Time                    &$t$     &$[L]/[u]$                             &$2$                 \hspace{-0.2cm}            &$\s$          \\
  Temperature             &$T$     &$[T]$                                 &$1$                 \hspace{-0.2cm}            &$\K$          \\
  Energy Density          &$e$     &$[\rho][u]^2$                         &$1$                 \hspace{-0.2cm}            &\,J$\mcube$   \\
  Magnetic Field          &$B$     &$\sqrt{\mu_0}[\rho]^{\frac{1}{2}}[u]$ &$1.12..\times10^{-3}$\hspace{-0.2cm}            &\,T $^{\dagger}$          \\
  Pressure                &$p$     &$[\rho][u]^2$                         &$1$                 \hspace{-0.2cm}            &\,Pa          \\
  \hline
    \end{tabular}}}
  $^{\dagger}\,\mu_0=4\upi\times10^{-7}$\,N\,A$^{-2}$
  \end{table}

These equations can be non-dimensionalised by dividing the variables with 
typical 
units, as detailed in Table~\ref{tab:units}. 

\section{Solution to background static equilibrium }\label{sect:soln}
In this Appendix we explicitly outline the solution to Eqs.~\eqref{eq:dpdr} 
and \eqref{eq:dpdz}.

\subsection{Basic quantities and derivatives}
Listed here are the form of the magnetic field components and the various
 derivatives
of the expressions which will be required in the calculations.
\begin{equation*}
  {\freply{B_r=-f\GO
  \frac{\upartial \BO}{\upartial z}
  -r\frac{\upartial \Bbz}{\upartial z},\quad
  B_z={\BO^2\GO}+2\Bbz}},
\end{equation*}
\begin{equation*}
\frac{\upartial f}{\upartial r}={\BO}{},\quad 
\frac{\upartial f}{\upartial z}=
{r}{}
  \frac{\upartial \BO}{\upartial z}
,\quad 
\frac{\upartial \GO}{\upartial f}=-\frac{2f\GO}{\fO^2},\quad 
\end{equation*}
\begin{equation*}
\frac{\upartial \GO}{\upartial r}= 
\frac{\upartial \GO}{\upartial f}\frac{\upartial f}{\upartial r}=
-\frac{2\BO f\GO}{\fO^2 },\quad 
\frac{\upartial \GO}{\upartial z}=
-\frac{2 f\GO r}{\fO^2}
  \frac{\upartial \BO}{\upartial z}.
\end{equation*}

\subsection{Magnetic pressure terms}
The magnetic pressure terms will integrate directly in Eq.~\eqref{eq:pbeq}
 and so we shall not require the derivatives. They are noted here for inclusion
in the final result.
\begin{eqnarray*}
\frac{|\vect{B}|^2}{2}&=&
\frac{B_r^2}{2}+\frac{B_z^2}{2}\\
&=&
{\freply{\frac{1}{2}
\left(
{f\GO}
  \frac{\upartial \BO}{\upartial z}
+r\frac{\upartial\Bbz}{\upartial z}
\right)^2
+\frac{1}{2}
\left({\BO^2\GO}
+2\Bbz
\right)^2}}. 
\end{eqnarray*}

\subsection{Magnetic tension force}

The components of the magnetic tension force are given by the general
 expressions by components $\hat{\vect{r}}$ and $\hat{\vect{z}}$, respectively,
\begin{equation}\label{eq:xtens}
B_r \frac{\upartial B_r}{\upartial r} 
+B_z\frac{\upartial B_r}{\upartial z},
\end{equation}
\begin{equation}\label{eq:ztens}
B_r\frac{\upartial B_z}{\upartial r} 
+B_z\frac{\upartial B_z}{\upartial z}. 
\end{equation}
We will require the derivatives in these expressions as follows: 
{\freply{\begin{eqnarray}\label{eq:dbxdx}
\frac{\upartial B_r}{\upartial r}
&=& 
-\left(
{\BO\GO}+{f}\frac{\upartial \GO}{\upartial r}
\right)
  \frac{\upartial \BO}{\upartial z}
-\frac{\upartial\Bbz}{\upartial z}
\nonumber\\
&=& 
{\BO\GO}
\left(
\frac{2f^2}{\fO^2}-1
\right)
  \frac{\upartial \BO}{\upartial z}
-\frac{\upartial\Bbz}{\upartial z}
\end{eqnarray}
\begin{eqnarray}\label{eq:dbxdz}
\frac{\upartial B_r}{\upartial z}
&=& 
-\left(
{\GO }\frac{\upartial f}{\upartial z}+{f}\frac{\upartial \GO}{\upartial z}
\right)
  \frac{\upartial \BO}{\upartial z}
\nonumber\\
&&- 
f\GO
\frac{\upartial}{\upartial z}
\left(
  \frac{\upartial \BO}{\upartial z}
\right)
-r\frac{\upartial^2\Bbz}{\upartial z^2}
\\
&=& 
{\GO r}
\left(
\frac{2f^2}{\fO^2}-1
\right)
  \frac{\upartial \BO}{\upartial z}^2
-
f\GO
  \frac{\upartial^2 \BO}{\upartial z^2}
-r\frac{\upartial^2\Bbz}{\upartial z^2}
\nonumber
\end{eqnarray}
\begin{eqnarray}\label{eq:dbzdx}
\frac{\upartial B_z}{\upartial r}
&=& 
{\BO^2}\frac{\upartial \GO}{\upartial r}
= -\frac{2\BO^3f\GO}{\fO^2},
\end{eqnarray}
\begin{eqnarray}\label{eq:dbzdz}
\frac{\upartial B_z}{\upartial z}
&=& 
{2\BO\GO}\frac{\upartial \BO}{\upartial z}
+{\BO^2}\frac{\upartial \GO}{\upartial z}
+2\frac{\upartial \Bbz}{\upartial z}
\nonumber
\\
&=&
{2\BO\GO}
\left(
1-\frac{f^2}{\fO^2}
\right)
  \frac{\upartial \BO}{\upartial z}
+2\frac{\upartial \Bbz}{\upartial z}
\end{eqnarray}
}}

{\freply{\subsection{Thermal pressure balancing magnetic field}}}

Having prescribed the magnetic field
we now seek to satisfy the pressure balance, first by 
solving Eq.~\eqref{eq:dpdr} 
for the ${\vect{r}}$-components.
The first term of the right-hand side below is magnetic pressure.
Subsequent terms yield the expression Eq.~\eqref{eq:xtens} by 
multiplying $B_r$ with \eqref{eq:dbxdx} and $B_z$ with \eqref{eq:dbxdz}.
{\freply{
{\footnotesize{
\begin{eqnarray*}
\frac{\upartial p_\hh}{\upartial r}
&=&
-\frac{\upartial }{\upartial r}\left(
\frac{|\vect{B}|^2}{2}
\right)
+\cancel{
{f\GO} 
\cdot
{\BO\GO}
\left(
\frac{2f^2}{\fO^2}-1
\right)
  \frac{\upartial \BO}{\upartial z}^2}
\nonumber \\
&&
-
f\GO
  \frac{\upartial \BO}{\upartial z}
\cdot
\frac{\upartial \Bbz}{\upartial z}
+
r\frac{\upartial \Bbz}{\upartial z}
\cdot
{\BO\GO}
\left(
\frac{2f^2}{\fO^2}-1
\right)
  \frac{\upartial \BO}{\upartial z}
\nonumber \\
&&
-{r\frac{\upartial \Bbz}{\upartial z}
\cdot
\frac{\upartial \Bbz}{\upartial z}}
-
\cancel{
\BO^2\GO\cdot
{\GO r}
\left(
\frac{2f^2}{\fO^2}-1
\right)
  \frac{\upartial \BO}{\upartial z}^2}
\nonumber \\
&&
+\BO^2\GO\cdot f\GO
  \frac{\upartial^2 \BO}{\upartial z^2}
-2\Bbz\cdot
{\GO r}
\left(
\frac{2f^2}{\fO^2}-1
\right)
  \frac{\upartial \BO}{\upartial z}^2
\nonumber \\
&&
+\BO^2\GO\cdot
r\frac{\upartial^2 \Bbz}{\upartial z^2}
+{2\Bbz\cdot
r\frac{\upartial^2 \Bbz}{\upartial z^2}}
+2\Bbz\cdot
f\GO 
  \frac{\upartial^2 \BO}{\upartial z^2}
\end{eqnarray*}
\begin{eqnarray*}
\ldots&=&
-\frac{\upartial }{\upartial r}\left(
\frac{|\vect{B}|^2}{2}
\right)
+
\frac{\upartial}{\upartial r}
\left(
\frac{2\Bbz f^2\GO}{\BO^2}
+
\frac{\Bbz\fO^2\GO}{\BO^2}
\right)
  \frac{\upartial \BO}{\upartial z}^2
\nonumber \\
&&
-
\frac{\Bbz\fO^2}{\BO}
\frac{\upartial\GO}{\upartial r}
  \frac{\upartial^2 \BO}{\upartial z^2}
-
\frac{\BO\fO^2}{4}
\frac{\upartial \GO^2}{\upartial r}
  \frac{\upartial^2 \BO}{\upartial z^2}
\nonumber \\
&&
-
\frac{\upartial}{\upartial r}
\left(
\frac{ f^2\GO}{\BO}
+
\cancel{
\frac{\fO^2\GO}{2\BO}}
\right)
\frac{\upartial \Bbz}{\upartial z}
  \frac{\upartial \BO}{\upartial z}
+
2r\Bbz\frac{\upartial^2 \Bbz}{\upartial z^2}
\nonumber \\
&&
+
\cancel{\frac{\fO^2}{2\BO}
\frac{\upartial\GO}{\upartial r}
\frac{\upartial \Bbz}{\upartial z}
  \frac{\upartial \BO}{\upartial z}}
-\frac{\fO^2}{2}
\frac{\upartial\GO}{\upartial r}
\frac{\upartial^2 \Bbz}{\upartial z^2}
-r\frac{\upartial \Bbz}{\upartial z}^2
\end{eqnarray*}
}}
\begin{eqnarray}\label{eq:phfull}
p_\hh
&=&
-\frac{|\vect{B}|^2}{2}
+
\left(
{\frac{2\Bbz f^2\GO}{\BO^2}}
+{\frac{\Bbz \fO^2\GO}{\BO^2}}
\right)
  \frac{\upartial \BO}{\upartial z}^2
 \\
&&
-
\frac{\Bbz\fO^2\GO}{\BO}
  \frac{\upartial^2 \BO}{\upartial z^2}
-\frac{\BO\fO^2\GO^2}{4}
  \frac{\upartial^2 \BO}{\upartial z^2}
-\frac{r^2}{2}\frac{\upartial \Bbz}{\upartial z}^2
\nonumber \\
&&
-
\frac{f^2\GO}{\BO}
\frac{\upartial \Bbz}{\upartial z}
  \frac{\upartial \BO}{\upartial z}
-
\frac{\fO^2\GO}{2}
\frac{\upartial^2 \Bbz}{\upartial z^2}
+r^2\Bbz\frac{\upartial^2 \Bbz}{\upartial z^2}.\nonumber
\end{eqnarray}
The solution is constrained by $p=p\vv+p\hh$ such that any 
constant of integration, a function of $z$, may be expressed within $p\vv$.  
Note that this solution can be simplified if our model can neglect the ambient 
magnetic field $\Bbz$, which outside the flux tube would result in
plasma-$\beta>1$ in the corona and the chromosphere. Then 
\begin{equation}\label{eq:pmred}
p_\hh
=
-\frac{|\vect{B}|^2}{2}
-\frac{\BO\fO^2\GO^2}{4}
  \frac{\upartial^2 \BO}{\upartial z^2}
\end{equation}
}} 
\subsection{Plasma density balancing magnetic field}

To determine $\rho\hh$
it is also necessary to integrate 
$\upartial p_\hh/\upartial z$ in Eq.~\eqref{eq:dpdz}.
For the magnetic tension terms of Eq.~\eqref{eq:ztens}, $B_r$ is 
multiplied with the expression \eqref{eq:dbzdx} and 
 $B_z$ with \eqref{eq:dbzdz}.  
\begin{eqnarray}\label{eq:adpdz}
\frac{\upartial p_\hh}{\upartial z}
&=&
\rho_\hh g
-\frac{\upartial }{\upartial z}\left(
\frac{|\vect{B}|^2}{2}
\right)
-
r\frac{\upartial \Bbz}{\upartial z}
\cdot
\frac{2\BO^3f\GO}{\fO^2}
\\\nonumber
&&
-
\cancel{f\GO
  \frac{\upartial \BO}{\upartial z}
\cdot
\frac{2\BO^3f\GO}{\fO^2}}
-\BO^2\GO\cdot
{\freply{2\frac{\upartial \Bbz}{\upartial z}}}
\\\nonumber
&&
-\BO^2\GO\cdot
2\BO\GO\left(
1-\cancel{\frac{f^2}{\fO^2}}
\right)
  \frac{\upartial \BO}{\upartial z}
\\\nonumber
&&
-{\freply{2\Bbz\cdot
2\BO\GO\left(
1-\frac{f^2}{\fO^2}
\right)
  \frac{\upartial \BO}{\upartial z}
+2\Bbz\cdot
2\frac{\upartial \Bbz}{\upartial z}}}
\end{eqnarray}
The solution to this must match that of Eq.~\eqref{eq:phfull}.
The match can be more easily identified if we add to this
the following list of terms, each  equating to zero:
{\freply{
{\footnotesize{
\begin{eqnarray*}
+
\frac{\upartial }{\upartial z}
\left(
{2 \GO r^2}\Bbz
  \frac{\upartial \BO}{\upartial z}^2
\right)
-\cancel{2}{\GO r^2
\frac{\upartial \Bbz}{\upartial z}
  \frac{\upartial \BO}{\upartial z}^2
}
-
{4 \GO r^2}\Bbz
  \frac{\upartial \BO}{\upartial z}
  \frac{\upartial^2 \BO}{\upartial z^2}
\nonumber \\
+\frac{4f \GO r^3}{\fO^2}\Bbz 
  \frac{\upartial \BO}{\upartial z}^3,
\end{eqnarray*}
\begin{eqnarray*}
+
\frac{\upartial }{\upartial z}
\left(
\frac{ \Bbz \fO^2\GO}{\BO^2}
  \frac{\upartial \BO}{\upartial z}^2
\right)
+\frac{2\Bbz  \GO r^2}{\BO} 
  \frac{\upartial \BO}{\upartial z}^3
-
\frac{ 2\Bbz \fO^2\GO}{\BO^2}
  \frac{\upartial \BO}{\upartial z}
  \frac{\upartial^2 \BO}{\upartial z^2}
\nonumber \\
-{
\frac{\fO^2\GO}{\BO^2}
\frac{\upartial \Bbz}{\upartial z}
  \frac{\upartial \BO}{\upartial z}^2
}^{}
+\frac{2\Bbz\fO^2 \GO}{\BO^3} 
  \frac{\upartial \BO}{\upartial z}^3,
\end{eqnarray*}
\begin{eqnarray*}
-
\frac{\upartial }{\upartial z}
\left(
\frac{\BO\fO^2\GO^2}{4}
  \frac{\upartial^2 \BO}{\upartial z^2}
\right)
+
\left(
\frac{\fO^2\GO^2}{4}
-
 f^2\GO^2 
\right)
  \frac{\upartial \BO}{\upartial z}
  \frac{\upartial^2 \BO}{\upartial z^2}
\nonumber \\
+
\frac{\BO\fO^2\GO^2}{4}
  \frac{\upartial^3 \BO}{\upartial z^3},
\end{eqnarray*}
\begin{eqnarray*}
-
\frac{\upartial }{\upartial z}
\left(
\frac{\Bbz \fO^2\GO}{\BO}
  \frac{\upartial^2 \BO}{\upartial z^2}
\right)
-
\left(
{\frac{ \fO^2\GO}{\BO^2}}
+
{{2 \GO r^2}}
\right)\Bbz
  \frac{\upartial \BO}{\upartial z}
  \frac{\upartial^2 \BO}{\upartial z^2}
\nonumber \\
+
{{\frac{\Bbz \fO^2\GO}{\BO}}
  \frac{\upartial^3 \BO}{\upartial z^3}}
+
{\frac{\fO^2\GO}{\BO}
\frac{\upartial \Bbz}{\upartial z}
  \frac{\upartial^2 \BO}{\upartial z^2}},
\end{eqnarray*}
\begin{eqnarray*}
-\frac{\upartial }{\upartial z}
\left(
{ f\GO r}{}
\frac{\upartial \Bbz}{\upartial z}
  \frac{\upartial \BO}{\upartial z}
\right)
+
{ f\GO r}{}
\frac{\upartial \Bbz}{\upartial z}
  \frac{\upartial^2 \BO}{\upartial z^2}
+
\bcancel{
{f\GO r}{}
\frac{\upartial^2 \Bbz}{\upartial z^2}
  \frac{\upartial \BO}{\upartial z}
}
\\
+
\left(
\cancel{{\GO r^2}{}}
-
\frac{2 f^2\GO r^2 }{\fO^2}
\right)
\frac{\upartial \Bbz}{\upartial z}
  \frac{\upartial \BO}{\upartial z}^2,
\end{eqnarray*}
\begin{eqnarray*}
-
\frac{\upartial }{\upartial z}
\left(
\frac{\fO^2\GO}{2}
\frac{\upartial^2 \Bbz}{\upartial z^2}
\right)
+
\frac{\fO^2\GO}{2}
\frac{\upartial^3 \Bbz}{\upartial z^3}
-
\bcancel{
{ f\GO r}{}
\frac{\upartial^2 \Bbz}{\upartial z^2}
  \frac{\upartial \BO}{\upartial z}
},
\end{eqnarray*}
\begin{eqnarray*}
+
\frac{\upartial }{\upartial z}
\left(
\Bbz r^2\frac{\upartial^2 \Bbz}{\upartial z^2}
-
\frac{ r^2}{2}
\frac{\upartial \Bbz}{\upartial z}^2
\right)
+
\Bbz r^2\frac{\upartial^3 \Bbz}{\upartial z^3}
.
\end{eqnarray*}
}}}}
If we filter out all of the derivative expressions, which can be integrated 
directly to return the same result as Eq.~\eqref{eq:phfull} any residual 
terms must
disappear and hence 
we require 
{\freply{
{\footnotesize{
\begin{eqnarray}\label{eq:initr}\label{eq:intdz}
\int \dd z
\rho_\hh g
&-&
\left[
\frac{2\BO^4\GO r^2}{\fO^2}
+{2\BO^2\GO}
-
4\Bbz
\right]
\frac{\upartial \Bbz}{\upartial z}
\nonumber \\
&+&
\frac{\fO^2\GO}{2}
\frac{\upartial^3 \Bbz}{\upartial z^3}
+
\Bbz r^2\frac{\upartial^3 \Bbz}{\upartial z^3}
-
2\BO^3\GO^2
  \frac{\upartial \BO}{\upartial z}
\nonumber \\
&-&
{{4\Bbz\BO\GO}}
\left[
1-\frac{f^2}{\fO^2}
\right]
  \frac{\upartial \BO}{\upartial z}
+\frac{ \fO^2\GO}{\BO}
\frac{\upartial \Bbz}{\upartial z}
  \frac{\upartial^2 \BO}{\upartial z^2}
\nonumber\\
&+&
{ f\GO r}{}
\frac{\upartial \Bbz}{\upartial z}
  \frac{\upartial^2 \BO}{\upartial z^2}
-
\frac{3\Bbz\fO^2\GO}{\BO^2}
  \frac{\upartial \BO}{\upartial z}
  \frac{\upartial^2 \BO}{\upartial z^2}
\nonumber \\
&+&
\left[
\frac{\fO^2\GO^2}{4}
-{f^2\GO^2}
-6
\Bbz
\GO r^2
\right]
  \frac{\upartial \BO}{\upartial z}
  \frac{\upartial^2 \BO}{\upartial z^2}
\nonumber \\
&-&
\left[
\frac{2 f^2\GO r^2}{\fO^2}
+
\frac{ \fO^2\GO}{\BO^2}
+
{ \GO r^2}
\right]
\frac{\upartial \Bbz}{\upartial z}
  \frac{\upartial \BO}{\upartial z}^2
\nonumber \\
&+&
\Bbz\GO\left[
\frac{ r^2}{\BO}
+\frac{2\fO^2 }{\BO^3}
+\frac{4 f r^3}{\fO^2}
\right]
  \frac{\upartial \BO}{\upartial z}^3
\nonumber \\
&+&
\frac{\BO\fO^2\GO^2}{4}
  \frac{\upartial^3 \BO}{\upartial z^3}
+
\frac{\Bbz\fO^2\GO}{\BO}
  \frac{\upartial^3 \BO}{\upartial z^3}
  = 0.
\end{eqnarray}
}}

\subsection{Divergence and pressure balance precision}

  In cylindrical polar coordinates the divergence of the magnetic field is 
  given by
  \begin{eqnarray}\label{eq:divB}
    \nabla\cdot\vect{B} &=& \frac{1}{r}\frac{\upartial }{\upartial r}(rB_r) + 
            \cancel{  \frac{1}{r}\frac{\upartial B_\phi}{\upartial \phi}} +
                           \frac{\upartial B_z}{\upartial z}
   \\
   \nonumber
    &=& -\frac{1}{r}\left[\frac{\upartial f}{\upartial z}\BO\GO + 
                          r\frac{\upartial^2 f}{\upartial r\upartial z}\BO\GO+
 r\frac{\upartial f}{\upartial z}\BO\frac{\upartial \GO}{\upartial f}\frac{\upartial f}{\upartial r}
      \right]
   \\
   \nonumber
    &+& \frac{\upartial f}{\upartial r}\frac{\upartial \BO}{\upartial z}\GO + 
                          \frac{\upartial^2 f}{\upartial r\upartial z}\BO\GO+
 \frac{\upartial f}{\upartial r}\BO\frac{\upartial \GO}{\upartial f}\frac{\upartial f}{\upartial z}
   \\
   \nonumber
  &-&\frac{1}{r}2r\frac{\upartial \Bbz}{z} + 2\frac{\upartial \Bbz}{z} = 0.
  \end{eqnarray}
  The resulting magnetic field
  configuration has been checked numerically with a mesh resolution
  $\delta x=10\km$ to verify
  \[
    \nabla\cdot\vect{B}=\frac{\upartial B_x}{\upartial x}
                       +\frac{\upartial B_y}{\upartial y}+
                        \frac{\upartial B_z}{\upartial z}=0.
  \]
  The resulting error scaled by the local strength of the field
  has mean of order $10^{-7}$, with peak of order $10^{-4}$.

  The horizontal pressure balance 
  \[
   \frac{\upartial p}{\upartial x} + 
   \frac{\upartial }{\upartial x}\frac{|\vect{B}|^2}{2}
   +B_x\frac{\upartial B_x}{\upartial x}
   +B_y\frac{\upartial B_x}{\upartial y}
   +B_z\frac{\upartial B_x}{\upartial z} = 0
  \]
  and vertical pressure balance
  \[
   \frac{\upartial p}{\upartial z} + 
   \frac{\upartial }{\upartial z}\frac{|\vect{B}|^2}{2}
   +B_x\frac{\upartial B_z}{\upartial x}
   +B_y\frac{\upartial B_z}{\upartial y}
   +B_z\frac{\upartial B_z}{\upartial z} - \rho g = 0
  \]
  have been verified numerically with $\delta x=10\km$ for the derived 
  thermal pressure, density
  and specified magnetic field configuration.
  For the horizontal pressure balance $\epsilon<10^{-13}$ and for the vertical
  mean relative error $\epsilon\simeq10^{-7}$ with peak of order $10^{-4}$.
  As $\delta x\rightarrow 0$ the relative error $\epsilon\rightarrow0$.
  }}  

  For these and we use 
  the same derivative scheme as applied in the 
  Versatile Advection Code \citep{Toth96} and the Sheffield Advanced
  Code for MHD \citep{SFE08}, which we plan to employ for future 
  simulations. 

\end{document}